# Mapping Interfacial Energetic Landscape in Organic Solar Cells Reveals Pathways to Reducing Nonradiative Losses

*Gaurab J. Thapa[1,2,†], Mihirsinh Chauhan[1, †], Jacob P. Mauthe[1], Daniel B. Dougherty[2*], Aram Amassian[1*]*

[1]*Department of Materials Science and Engineering and Organic and Carbon Electronics Laboratories (ORaCEL), North Carolina State University, Raleigh, NC, 27695, USA.*

[2]*Department of Physics and Organic and Carbon Electronics Laboratories (ORaCEL), North Carolina State University, Raleigh, NC, 27695 USA.*

[†] *These Authors contributed equally*

*Authors to whom correspondence should be addressed: dbdoughe@ncsu.edu and aamassi@ncsu.edu

KEYWORDS: Organic photovoltaics, Organic solar cells, Bulk heterojunction, Energetic disorder, Charge transfer state, Nonradiative recombination, Energetic landscape, Donor-acceptor interface, Scanning tunneling microscopy/spectroscopy, Non-fullerene acceptors, sensitive EQE, Marcus theory.

**Summary**

Bulk heterojunction (BHJ) organic solar cells have made remarkable inroads towards 20% power conversion efficiency, yet nonradiative recombination losses ($\Delta V_{nr}$) remain the highest as compared to silicon and perovskite photovoltaics. Interfaces buried within BHJ blends hold the key to recombination losses but insights into their energetic landscape underpinning charge transfer (CT) states and their disorder remain elusive. Here, we reveal the energetic landscape and CT state manifold of modern BHJs with both spatial and energetic resolutions and link the offset between singlet ($E_{S1}$) and CT energy ($E_{S1-CT}$) and interfacial energetic disorder with $\Delta V_{nr}$. We do so by locally mapping the energy distributions of modern PM6-based BHJs with IT4F, Y6 and PC$_{71}$BM acceptors and combine it, for the first time, with sensitive EQE measurements, to visualize and quantify donor (D) and acceptor (A) energetics at interfaces and associated them with CT states within a modified Marcus framework. A key new ability is the identification of the specific BHJ interfaces associated with the CT manifold, including where the lowest energy CT states reside. Moreover, we quantify energy levels and electronic disorders directly at these and other interfaces and connect these contributions to the energy losses. We delineate the influences of $S_1$ to CT offset and interfacial energetic disorder on the $\Delta V_{nr}$ across multiple morphologically varied BHJs. Our results clearly show both factors influencing energy losses and that changing the interfacial disorder affects non-radiative voltage losses in systems with comparable $S_1$ to CT offset. We demonstrate that PM6:Y6 can achieve low $\Delta V_{nr}$ by forming a nominally sharp D/A interface with exceptionally low interfacial disorder via judicious processing combined with a low $S_1$ to CT offset. This provides a design rule to minimize $\Delta V_{nr}$ for modern NFAs: sharp D/A interfaces with low $S_1$ to CT offset exhibiting minimal interfacial disorder.

**1.Introduction:**

With the advent of non-fullerene acceptors (NFAs), organic photovoltaics (OPVs) have made remarkable progress[1,2]. However, it is necessary to gain a deeper understanding of the underlying loss mechanisms limiting the operational performance of these devices.

In particular, significant nonradiative charge recombination occurs in OPVs as compared to other thin film photovoltaic technologies[3–5]. An overwhelming body of work now shows the influence of charge transfer (CT) state properties on the charge generation and recombination processes for both fullerenes and NFAs based OPVs[5–10]. In addition, the energy gap law[11] suggests that the resulting non-radiative voltage loss ($\Delta V_{nr}$), which depends on the electroluminescence efficiency[12], decreases with increasing CT state energy, and the offset between the singlet ($E_{S1}$) and lowest lying CT state ($E_{CT}$) energies leads to a lower $\Delta V_{nr}$[13]. Interestingly, OPVs based on some NFAs show an opposite trend to the energy gap law due to their negligible ($E_{S1} - E_{CT}$) offsets and associated hybridization[5]. In modern NFA based OPVs, it has been shown that the $\Delta V_{nr}$ pathways proceed through formation of triplet excitons[4]. Overall, these findings signify the importance of accurately pinpointing and identifying the nature of the interfaces and underlying disorder of the CT states in a BHJ to fully specify the mechanism of non-radiative voltage losses in OPVs.

The CT state electronic properties at donor (D) /acceptor (A) interfaces include the disorder present in active materials in proximity of D/A interfaces[8]. A variety of intra and intermolecular interactions and a rich conformational diversity can give rise to static electronic disorder, as well as multiple CT excited states within a bulk heterojunction (BHJ)[8,14]. In addition, the temperature-induced molecular motion broadens CT state distributions and is referred to as dynamic disorder[15]. Both static and dynamic disorder lead to fluctuations in the electronic states including the ionization potential (IP) and electron affinity (EA) of the OPV materials[9]. Thus, by measuring the statistical distribution of the electronic levels, we can quantify the total disorder ($\sigma_T$) as the sum in quadrature of the static and dynamic counterparts, i.e. $(\sigma_T)^2 = ((\sigma_s)^2 + (\sigma_d)^2)$[9].

Depending on the type of interface formed between D and A materials and the associated energetic landscape, CT states can exhibit a broad range of statistical distributions of energies[8,14]. It has been generally established that the interfaces between ordered D and A materials give rise to low energy CT states ($E_{D/A}$), thus making these the dominant "functional" interfaces of the solar cell, i.e., where charges are generated and recombine at open circuit voltage ($V_{oc}$) conditions[6,16]. In contrast, interactions between disordered D and A materials in the mixed (M) phase give rise to higher energy CT states ($E_{M/M}$), and therefore less likely to participate in charge generation and recombination at $V_{oc}$ in the presence of other low energy CT states[16,17].

Techniques designed to investigate the nature and properties of CT states, like sensitive-external quantum efficiency (s-EQE) and emission spectra[18] provide information about the lowest lying CT state properties as well as their overall energetic disorder without providing any specificity about the nature of the interfaces or molecular interactions, nor about the specific D and A electronic disorder contributions. A recent body of work has focused on measuring the energetic disorder of bulk materials, including nanoscale domains within the BHJ using spectroscopic techniques and x-ray scattering techniques[19,20]. However, it is unclear whether the degree of energetic or structural order within bulk phases or even nanoscale domains will translate into energetic disorder at the interfaces that dictate the CT state disorder linked with nonradiative voltage losses in OPV devices[21].

A precise understanding of the nature of CT states and the local energetic landscape of modern three-phase BHJ morphologies with multiple types of CT states[14] is essential further reducing nonradiative recombination in OPVs as we move beyond 20% PCE. However, existing spectroscopic techniques spatially average over the complex landscape of different interfaces, making it challenging to identify which one is the active CT state and what energetic disorder is present. Existing techniques for measuring the frontier molecular energy levels include cyclic voltammetry (CV), ultraviolet photoelectron spectroscopy/inverse photoelectron spectroscopy (UPS/IPES) and energy-resolved impedance spectroscopy (ER-EIS)[5,19,22,23]. Among these, UPS/IPES measurements provide best known correlations with device operational figures of merit[22]. However, there are some limitations with the IPES measurements due to the chance of film damage by high electron energy irradiation[24], poor accuracy (±0.35 eV)[22], and sample charging issues[25]. In addition, UPS/IPES is usually performed on neat films rather than the actual BHJ and neat material properties are assumed to be unchanged in simplified device models. Furthermore, it cannot resolve different phases, nor can it assign energetic distributions to phases or interfaces. This can cause ambiguity as the energetic distributions of different phases and interfaces within a BHJ are highly dependent on the thermodynamics of materials and their blends and kinetics of processing, all of which determine the BHJ morphology[26–29]. To address this, Salleo et al. proposed a framework to establish a direct relationship between the NFA blend morphology and structural (dis)order and its influence on the CT state energy distribution by varying the composition of the blend[20]. This method has made important but qualitative progress towards helping pinpoint which morphological features influence the CT state disorder and are in turn responsible for nonradiative recombination. However, it relies on measures of structural disorder taken from bulk domains as a surrogate for electronic disorder at functional interfaces.

In this work, we directly and locally map the energetic distribution of electronic states at modern BHJ interfaces. We use scanning tunneling microscopy/spectroscopy (STM/S) to evaluate interfaces accessible through the top surface of BHJ films and combine the energetic landscape analysis of CS states with mean-field s-EQE analysis of CT states on similar BHJ films within a modified Marcus framework, achieving excellent agreement. Our framework allows one to visualize, identify, and characterize D-rich, A-rich, and M phases within the BHJ and all associated D/A, D/M, M/A and M/M interfaces and CT states by combining local conductance mapping and local tunneling spectroscopy. It also provides electronic distributions in local ionization potential (IP) and electronic affinity (EA) levels at different interfaces to guide understanding of excitonic CT states associated charge generation and recombination. Our study is conducted across PM6-based blends with IT4F, Y6 and $PC_{71}BM$ acceptors yielding significantly different morphologies and nonradiative voltage losses ranging from 0.23 V to 0.4 V. We reveal how the interfacial energetic disorder of PM6 and the acceptor can differ from the disorder deeper within the BHJ domains. These observations justify the necessity for direct and local measurements on BHJ interfaces when seeking to quantify interfacial electronic disorder associated with various CT states. Utilizing STM/S, for the first time, we directly identify different interfaces in BHJs at the nanoscale and by combining it with s-EQE, we obtain electronic disorder at these interfaces and the ($E_{S1} - E_{CT}$) offset and relate it to the open-circuit voltage losses using the combined STM/S and s-EQE methodology. Our study reveals that one of the reasons for PM6:Y6 exhibiting lower nonradiative recombination losses than the other materials is due to the combination of low the ($E_{S1} - E_{CT}$) offset and its ability to achieve enhanced local interfacial electronic properties at its "sharp" D/A interface *via* judicious processing. Through our framework, we demonstrate this is achieved by significantly reducing the interfacial electronic disorder of PM6 and Y6 and the associated CT state disorder at the sharp D/A interface.

In addition, we evaluate the influence of both the ($E_{S1} - E_{CT}$) offset and the interfacial electronic disorder upon the non-radiative voltage loss within the same D/A system and across multiple morphologically varied BHJs. We confirm that when the ($E_{S1} - E_{CT}$) offset is fixed, the interfacial disorder significantly influences the non-radiative voltage loss. Our study shows a pathway to minimizing energy losses through selection of materials with low ($E_{S1} - E_{CT}$) offset and optimal miscibility that achieve reduced electronic disorder of both D and A components at functional interfaces.

## 2. Results and Discussion:

### 2.a The Combined STM/S and s-EQE Methodology

In this section we describe a new method of combining s-EQE analysis in a modified Marcus model with local STM/S measurements of interfacial CS state energetics and disorder. The combination allows us to extract essential information about the BHJ energetic landscape and CT state manifold. Experimental access to this information has been limited in the past, but this new approach enables the association of identified CT states with visualized interfaces. It also provides independent cross-validation of disorder measurements via the apparent Urbach analysis[30]. In addition to successfully improving the fitting accuracy and holistically quantifying the energetic landscape, we further aim to visualize the BHJ and differentiate the behavior of D and A materials within domains compared to interfaces and uniquely identify the functional interface most likely responsible for charge recombination at $V_{oc}$. Importantly, STM/S provides information about the neat D, A materials, mixed phases, and interfacial regions within the BHJ (**Figure 1(A)**) which is not possible from the combination of s-EQE and EL measurements with structural measurements as these latter macroscopic measurements probe the entire BHJ at once.

STM/S has been successfully used to probe the electronic structure of organic semiconductors and lateral heterojunctions exposed at the top surface of the BHJ with sub-nanometer spatial resolution[31–37]. In particular, the differential tunneling conductance (dI/dV) with respect to the tip-sample bias (V) probes the local density of states (LDOS) and can be used to spatially resolve electronic transport states. This approach (**Figure S1**) has been used to investigate the electronic structure of BHJs experimentally[34,35] and has been successfully applied in the past to organic semiconductor heterojunctions in the context of photovoltaics[33,38,39]. The IP and EA onsets obtained from the averaged dI/dV spectra using STM/S on neat materials in our study are in good quantitative agreement with the UPS/IPES on neat materials **(Figure S2)**. Within PM6: acceptor blends, STM/S mapping and spectroscopy allows the tip to probe D-, A-rich and mixed (M) phases in a typical three-phase BHJ. These are visualized as red, blue, and green regions, respectively, for the remainder of this study (details of the technique are explained in **Figure S3**). The IP and EA levels of the materials in the different domains can be probed locally and averaged over several dozen dI/dV spectra obtained from multiple dI/dV maps (**Figure S4**) to acquire statistics from a large area of interest. Additionally, interfacial regions, such as D/A, D/M, M/A and M/M can be visually located in dI/dV maps (**SI section 3**) and laterally probed with STS measurements to quantify the evolution of the energetic landscape from one nanoscale domain to the other. The dI/dV spectra collected exclusively within interfacial regions will have characteristics associated with the material constituents exhibiting the highest IP and lowest EA, therefore these will correspond to the donor's IP ($IP_D$) and the acceptor's EA ($EA_A$), respectively. Statistical averaging of dI/dV spectra targeting the nominally sharp interfacial regions versus the D- and A-rich domains, referred to as bulk in this context, enables clear differentiation between their respective energetic landscapes. By "sharp" interfaces, we refer to the spatial extent of the transition region between donor-rich and acceptor-rich domains as directly observed by STM imaging and local spectroscopy (**Figure S5**). dI/dV spectra acquired across sharp interfaces also show changes in the IP and EA levels similar to previous expectations[23,40]. Here, it is noteworthy that the scale bar on these images suggests that these interfaces are on the order of 1 nm or less which cannot be accessed by any other conventional methods (UPS/IPES)[41] and motivates our use of the word "sharp". These frontier levels may actively take part in charge generation via the formation of charge transfer (CT) states and their energetics may be used to compute the CT state energy and disorder with the appropriate framework[9,42].

In addition, we statistically analyze the energetic *differences* between the relevant IP and EA distribution (i: D/A, D/M, M/A, M) to retrieve the associated average energy difference ($E_i$) and electronic disorder parameter ($\sigma_T^i$). These energetic difference distributions of IP and EA onsets give direct access to the charge separated (CS) state distribution and have not been evaluated in prior STM/S work on BHJ's. Importantly, we will use this distribution of CS states as a surrogate measurement for CT states ($E_{CS} \sim E_{CT}$) and as direct input to s-EQE measurements. The methodology of combining the STM/S and s-EQE measurements is schematically illustrated in **Figure 1(B)**. This procedure leverages the negligible CT exciton binding energy[43] in modern acceptors and the relatively minor impact of interfacial electrostatic interaction disorder on the net electronic disorder[9].

Spatially averaged dI/dV spectra obtained from the interfacial region and various phases within the BHJ for PM6:IT4F (**Figure 1 (C)**) show the differences in the IP and EA onsets, especially for the interfacial region and the M phase where the former has reduced EA onset compared to the latter. The IP and EA onsets can be compiled from the individual dI/dV spectrum to construct a histogram representing the energetic disorder for the region of interest. The normalized histograms of IP and EA onsets obtained for different interfaces/phases for the PM6:IT4F system (**Figure 1(D)**) reveal the extent to which the same material can exhibit significantly different energetic landscapes within the same BHJ, depending on the domain and/or interface where the material is found **(Figure S6 and Table 1)**. A weighted Gaussian fit is performed to each of these histograms **(Figure S7 and Table 2)** to extract the average energy difference ($E_i$) and electronic disorder parameter ($\sigma_T^i$).

A key insight from our new combined methodology is that among the CS/CT distributions at the different phases/interfaces (**Figure 1(E)**), it is the "sharp" interfaces that have the lowest energy for PM6:IT4F with M/A slightly higher than D/A while D/M and M/M interfaces have comparatively higher energies (**Table 2**). The comparison between static disorder of both D/A and M/A interfaces obtained using our combined methodology to that obtained by the "apparent" Urbach energy analysis reveals the nature of the functional interface, as will be discussed below. Notably, the "apparent" Urbach energy analysis to estimate the static disorder[30] is different than the traditional Urbach energy analysis sometimes used to fit absorption edges[44].

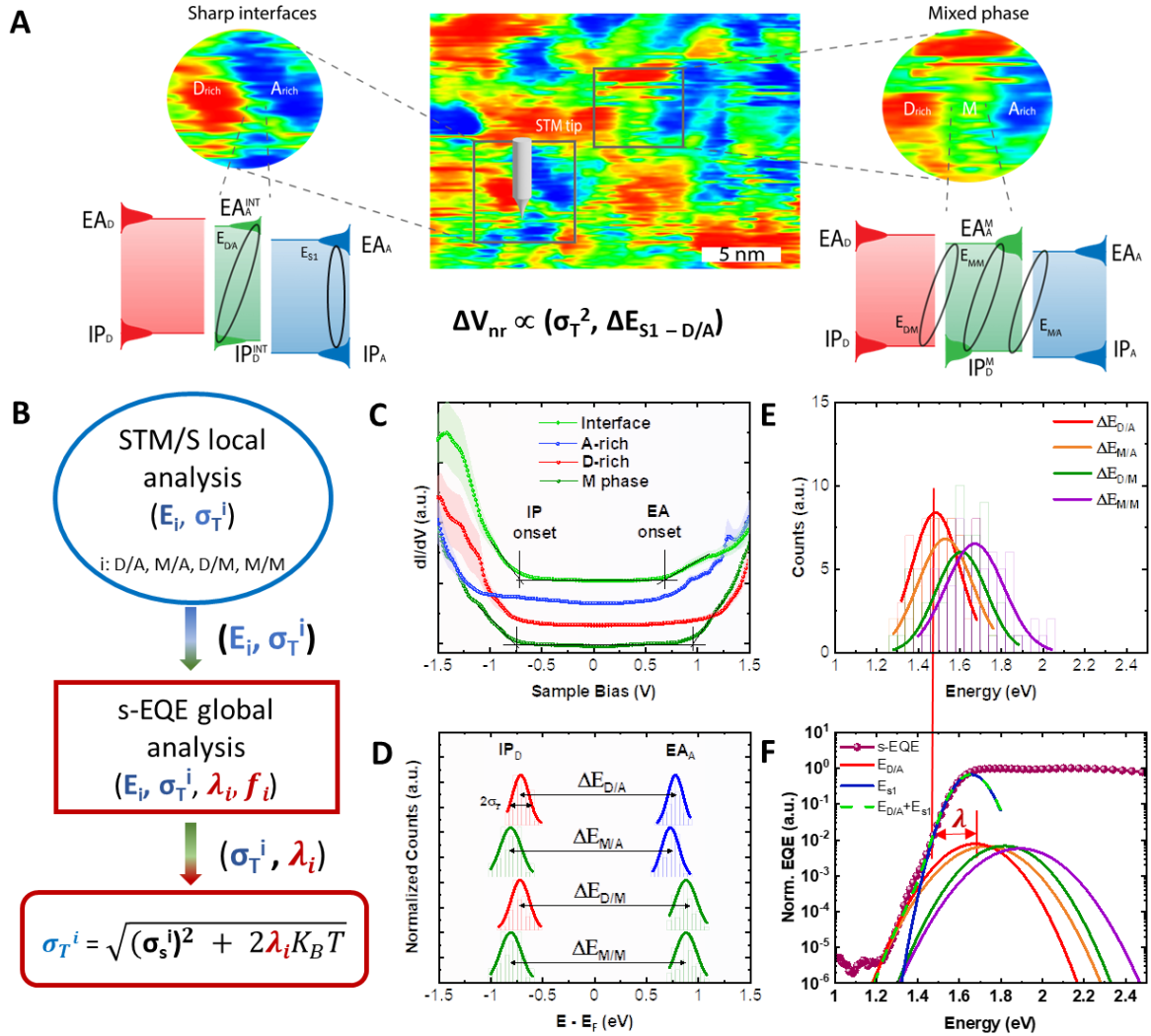

**Figure 1: Combined STM/S and s-EQE framework to reveal the CT states manifold.**

(A) STM/S dI/dV map of PM6:IT4F three-phase BHJ morphology with sharp interfaces between D and A and the mixed (M) phase. Spectroscopic measurements across these interfaces in combination with s-EQE allow to measure the complete CT states manifold of the BHJ, measure the $E_{S1-CT}$ offset and total disorder of the CT states responsible for non-radiative voltage losses in OPV devices. Schematic illustrations of the energetic diagrams of sharp interfaces and mixed phase. (B) Schematic illustration of the combined STM/S and s-EQE framework's analytical workflow allowing to obtain the CT states manifold of modern BHJs, energetic landscape, relevant offsets, and identify the functional interface. (C) Averaged dI/dV spectra obtained from the STS measurements from interfacial region and different phases within the PM6:IT4F BHJ (D) Normalized electronic distribution of donor IP and acceptor EA onsets measured in the interfacial region and different phases within the PM6:IT4F BHJ. (E) Electronic distribution obtained from the differences in the IP and EA onsets of sharp interfaces (D/A), M phase and A-rich phase (M/A), D-rich phase and M phase (D/M) and M phase (M/M) respectively for PM6:IT4F BHJ obtained from the STM/S analysis. (F) s-EQE from devices using PM6:IT4F as an active layer with the CT states fit corresponding to the different interfaces in PM6:IT4F BHJ using the electronic distribution obtained from the STM/S analysis.

The combined analytical framework that uses STM/S data to identify which specific interface plays a role in charge recombination processes is elaborated in detail here and in the **SI section 5**. The s-EQE analysis of PM6:IT4F uses direct inputs from the STM/S outputs of various CT state energies and disorders ($E_i$, $\sigma_T^i$, respectively) using a modified Marcus framework as expressed in equation (1)[9,45]:

$$EQE = \frac{f_i}{E\sqrt{2\pi(\sigma_T^i)^2}} exp\left(-\frac{(E_i + \lambda_i - E)^2}{2(\sigma_T^i)^2}\right) \qquad (1)$$

where $\lambda_i$ is the reorganization energy, $f_i$ is the oscillator strength of the i[th] CT state, and E is the energy. The total disorder ($\sigma_T^i$) is the sum of the static ($\sigma_s^i$) and dynamic disorder ($\sigma_d^i$) and related to the reorganization energy as,

$$(\sigma_T^i)^2 = (\sigma_s^i)^2 + (\sigma_d^i)^2; \; (\sigma_d^i)^2 = 2\lambda_i k_B T \qquad (2)$$

where $k_B$ is the Boltzmann constant and T is the absolute temperature. Except for the oscillator strength and reorganization energy, all the parameters in this expression can be accurately estimated from the CS state proxies obtained in our STS measurements. By combining experimental s-EQE measurements, we identify which interfaces visualized by STM/S host the CT states measured and which ones are most likely to be active in charge recombination at $V_{oc}$ based on lowest to highest energy and comparison with independent apparent Urbach energy analysis.

Assessing the CT states in OPVs has been performed by applying the framework of Marcus theory on experimental data obtained using the s-EQE or Fourier-transform photocurrent spectroscopy (FTPS) and electroluminescence (EL) spectra[18]. This method sheds light on the influence of CT states on the photovoltaic device performance and performed well for the fullerene and some NFAs based OPVs[46]. However, Y6 based OPVs showed significant absorption from the Y6 singlet dominating the s-EQE[47,48], where the CT state absorption shoulder becomes indistinguishable due to hybridization between the CT state and the singlet exciton. Moreover, the s-EQE provides information about the lowest energy CT states by doing a Marcus theory fit that includes three free parameters ($E_{CT}$, $\lambda$, f) and can cause ambiguity in fitting the s-EQE. Further, this theory does not consider static disorder, prevalent in organic semiconductors[49,50]. By modifying the theory to include the static disorder, the free parameter increases to $E_{CT}$, $\lambda$, f, and $\sigma_T$, which makes the reliable fitting more challenging.

Using the parameters appropriate to D/A and M/A interfaces provides a better fit to the s-EQE data as compared to the D/M and M/M interfaces (**Figure 1(F)**) showing that these interfaces primarily take part in charge recombination and thereby determine the $V_{oc}$. We obtained the static disorder ($\sigma_s^{D/A}$ = 58 ± 14 meV) of the CT state associated with the sharp interface (D/A) and M/A interface ($\sigma_s^{M/A}$ =79 ± 16 meV) using equation (2). To decide which of these interfaces play a major role in charge recombination processes in OSCs, we measured the static disorder using the apparent Urbach energy analysis following the model introduced by Kaiser et al.[30]. The quantitative agreement between the static disorder value obtained from the combined STM/S and sEQE analysis (**Figure S8**) for the D/A interface (58 ± 14 meV), unlike the M/A interface (79 ± 16 meV), and the apparent Urbach energy analysis (51 ± 2 meV) (**Figure S9**) is strong evidence that D/A play the dominant role in charge recombination processes and that the sharp D/A interface, as imaged by STM/S, controls the $V_{OC}$ and therefore the overall voltage loss in PM6:IT4F OPV devices.

## 2.b Application of the Combined Methodology to Blends with Different Miscibility

To generalize the viability of this methodology, we chose different acceptors to blend with PM6 resulting in diverse morphologies due to the thermodynamic and kinetic processes [26,27]. We evaluated energetic landscapes in other blend systems PM6:Y6 (**Figure S10 & S11**) and PM6:PC$_{71}$BM (**Figure S12 & S13**) categorized as hypomiscible and hypermiscible respectively mixed with the same PM6 donor. Hypomiscible and hypermiscible blend systems are expected to yield three-phase and (one)two-phase morphologies, respectively[26,51]. As expected, PM6:Y6 exhibits a three-phase morphology like PM6:IT4F, whereas PM6:PC$_{71}$BM is dominated by a two-phase morphology consisting of the M phase (green) and an A-rich mixed phase. This is definitively confirmed by local dI/dV point spectra (**Figure S14**).

To establish comparability with devices, we fabricated OPV devices of thickness used in STM/S measurements (~20 nm) (**SI section 8**). J-V and s-EQE measurements of thin and thick devices reveal differences that are expected given the significant differences in active layer thickness and optical interference effects, which have also been shown to impact the $V_{oc}$ and $V_{nr}$ losses **(Table S6)**[52]. Identical results between thin and thick devices can therefore not be expected and differences are within expectation based on prior studies. Importantly, we have compared the static disorder via apparent Urbach energy of the thin and thick BHJs **(Table S6)** and find that they are identical within error bars. Moreover, we performed AFM measurements to probe the topography of thick and thin films of these blend systems. We find that the surface topography of the thick and thin samples shows similar fibril structure suggesting that the underlying surface morphologies are similar in the thickness ranges we study with STM/S and in the devices (**Figure S17**).

Comparative views of the dI/dV maps of PM6:IT4F, PM6:Y6 and PM6:PC$_{71}$BM (**Figure 2 (A, E, I)**) reveal the spatial variations in local conductance of the respective BHJ morphologies. Comparison between the dI/dV spectra from the D- and A-rich phases within the BHJ and the neat D and A films highlight noticeable energetic shifts in IP and EA levels of the miscible PM6:PC$_{71}$BM blend (**Figure S12**). By contrast, they are far more similar for the immiscible PM6:IT4F (**Figure S6**) and partially immiscible PM6:Y6 (**Figure S10**) [26,51]. Here, we note that the distributions of the IP and EA levels in the blends are broader than that in the neat materials. We refer to these domains as D-rich and A-rich respectively, indicating that they are most likely not pure in composition. Moreover, the structural disorder in the enriched domains is likely to be at least somewhat different than the corresponding neat material and this will lead to corresponding differences in electronic disorder.

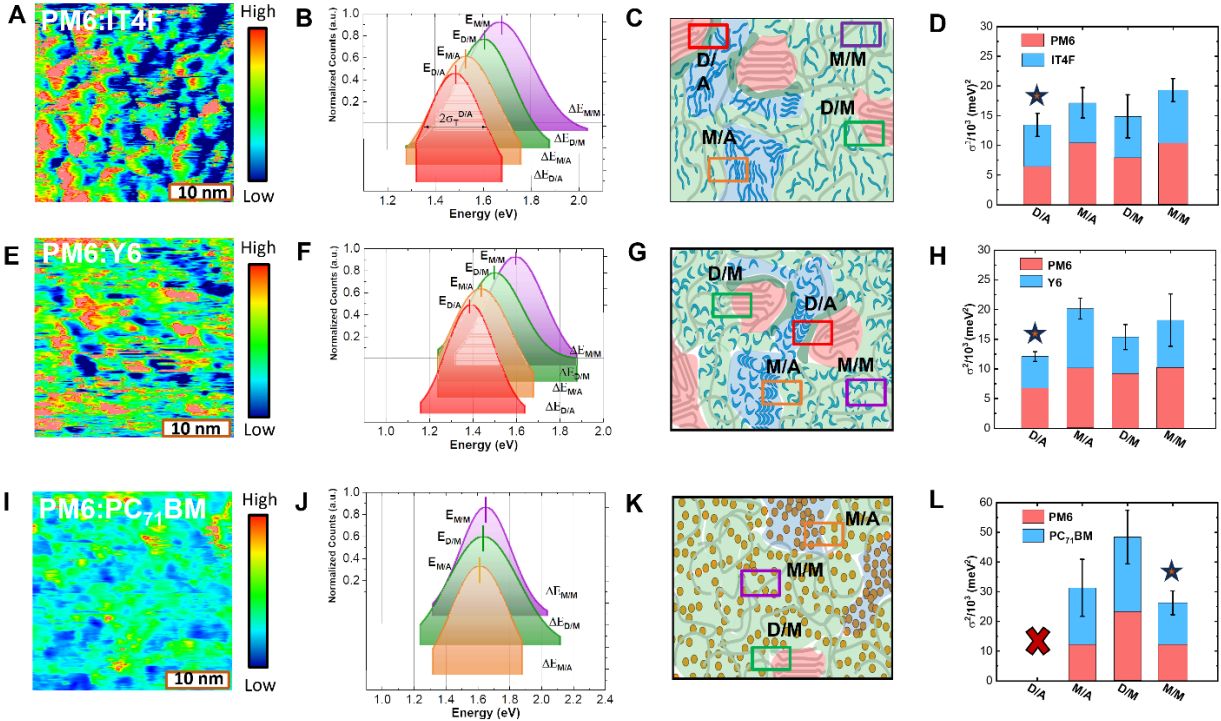

**Figure 2: STM/S analysis for PM6 based BHJ**

dI/dV maps obtained from STM/S measurements on a approx. 20 nm thin layer of (A) PM6:IT4F, (E) PM6:Y6 and (I) PM6:PC$_{71}$BM BHJs spin-coated on Au(111) surface and measured under ultrahigh vacuum (UHV) (5 × 10$^{-10}$ Torr) conditions and (-1V, 500 pA) tunneling setpoints. Gaussian fits to the distribution of energetic differences obtained from the donor IP and acceptor EA from sharp interfaces (D/A), interfaces between A-rich phase and M phase (M/A), interfaces between D-rich phase and M phase (D/M) and M phase (M) in (B) PM6:IT4F, (F) PM6:Y6 and (J) PM6:PC$_{71}$BM BHJs respectively. These fits were made to conductance onset histograms from 50-60 dI/dV spectra acquired from different spatial positions in the blend film corresponding to the PM6, A-rich and M phases using 0.04 eV bins. Schematic illustration of morphology for (C) PM6:IT4F, (G) PM6:Y6 and (K) PM6:PC$_{71}$BM highlighting the different interfaces in the respective BHJs. Squared Gaussian widths obtained from the distribution of different interfaces for (D) PM6:IT4F (H) PM6:Y6 and (L) PM6:PC$_{71}$BM along with the D and A contribution to the total squared width.

Donor and acceptor molecules coexist in the M phase, which dominates for the hypermiscible PM6:PC$_{71}$BM blend. The observed energetic shift of the donor due to mixing with fullerene acceptor is consistent with prior literature reports based on CV[53–55] and ER-EIS[56]. Further, UV-Vis spectroscopic analysis also suggests low aggregation of PM6-rich phase in PM6:PC$_{71}$BM as compared to neat PM6 (**Figure S18**). Our STM/S measurements of the M phase **(Table 2)** show no significant differences in the energy level of any of the possibly different interfaces (i.e. M/A, D/M and M/M) and a strongly disordered but relatively spatially uniform energetic landscape.

By contrast, the process of phase separation in hypomiscible systems leads to more spatial diversity in the energetic landscape. In addition to enriched domains and mixed domains, we observe sharp D/A interfaces with low-lying CT states in the PM6:IT4F and PM6:Y6 blends. **Figures 2(B, F and J)** show the best fit Gaussian distributions obtained from the energetic difference of the IPs and EAs of the D and A materials at different interfaces for PM6:IT4F **(Figure S7),** PM6:Y6 **(Figure S11),** and PM6:PC$_{71}$BM **(Figure S13)**, respectively. Like PM6:IT4F, the D/A and M/A interfaces provide a good fit to the tail region of the s-EQE measurement for the PM6:Y6(**Figure S20**). However, comparison of the static disorder obtained from the combined methodology for D/A (55 ± 9 meV) and M/A (89 ± 26 meV) interfaces with that obtained from the apparent Urbach analysis (48 ± 0.5 meV) shows that D/A interfaces plays a major role in charge recombination.

In contrast, the absence of sharp interfaces between D- and A-rich domains in PM6:PC$_{71}$BM is consistent with this system being hypermiscible. **Figures 2(C, G and K)** depict the underlying morphology inferred from our analysis to date for the three systems and indicate the CT states in PM6:PC$_{71}$BM are high energy in part due to the molecular mixing of D and A molecules across the BHJ. Quantitative analysis of static disorder from our framework combining STM/S and s-EQE analyses indicates that the functional interface at V$_{oc}$ for PM6:PC$_{71}$BM is the M phase itself (**Figure S20**). The squared

Gaussian width from the distributions associated to the various interfaces are represented in terms of contributions of the D and A respectively in (**Figures 2(D, H and L)**) and are summarized in **Table 2**.

The D and A contributions determined by STM/S help us evaluate the contribution of each constituent to the local disorder of the various interfaces or mixed phases and is plotted as squared Gaussian width ($\sigma_T^2/10^3$ (meV)$^2$) since the variance of donor IP and acceptor EA distribution adds up linearly to provide the total Gaussian variance. Interestingly, the interfaces between the D-rich and M phase in PM6:IT4F exhibit similar total disorder as the sharp interfaces but have noticeably higher energy than the sharp PM6/IT4F interface (**Table 2**) and therefore do not take part in charge recombination at $V_{oc}$. In contrast, the highly miscible nature of PM6 and PC$_{71}$BM gives rise to a lack of sharp interfaces. The major contribution to disorder in this blend appears to be from two factors: The inherent disorder in PC$_{71}$BM compared to modern NFAs[46,57], as well as its hypermiscibility with PM6[51] which disrupts ordering.

**Table 1: STM/S total disorder analysis for PM6 based BHJ**

Peak position and standard deviation obtained from the statistically weighted Gaussian fit to the histograms of the different phases in the studied PM6 based BHJ.

| | | IP | | EA | |
|---|---|---|---|---|---|
| **PM6:IT4F** | | Peak position (eV) | St. deviation (meV) | Peak position (eV) | St. deviation (meV) |
| | Neat PM6 | -0.732 ± 0.016 | 87 ± 13 | 1.184 ± 0.008 | 69 ± 6 |
| | PM6-rich | -0.717 ± 0.011 | 89 ± 10 | 1.207 ± 0.005 | 86 ± 4 |
| | Neat IT4F | -1.17 ± 0.012 | 68 ± 9 | 0.723 ± 0.008 | 69 ± 7 |
| | IT4F-rich | -1.165 ± 0.010 | 111 ± 13 | 0.729 ± 0.015 | 85 ± 13 |
| | Sharp Interface | -0.711 ± 0.014 | 80 ± 12 | 0.778 ± 0.008 | 74 ± 7 |
| | M phase | -0.808 ± 0.016 | 102 ± 15 | 0.88 ± 0.020 | 102 ± 19 |
| **PM6:Y6** | Neat PM6 | -0.73 ± 0.017 | 114 ± 17 | 1.206 ± 0.029 | 122 ± 32 |
| | PM6-rich | -0.707 ± 0.015 | 96 ± 13 | 1.227 ± 0.013 | 72 ± 12 |
| | Neat Y6 | -1.07 ± 0.007 | 82 ± 6 | 0.672 ± 0.008 | 85 ± 7 |
| | Y6-rich | -1.082 ± 0.026 | 106 ± 25 | 0.727 ± 0.012 | 97 ± 12 |
| | Interface | -0.661 ± 0.010 | 82 ± 8 | 0.736 ± 0.019 | 92 ± 18 |
| | M phase | -0.731 ± 0.011 | 101 ± 9 | 0.817 ± 0.063 | 192 ± 88 |
| **PM6:PC$_{71}$BM** | Neat PM6 | -0.732 ± 0.016 | 87 ± 13 | 1.184 ± 0.008 | 69 ± 6 |
| | PM6-rich | -0.833 ± 0.032 | 153 ± 34 | 0.885 ± 0.043 | 171 ± 51 |
| | Neat PC$_{71}$BM | -1.26 ± 0.011 | 89 ± 10 | 0.842 ± 0.016 | 94 ± 13 |
| | PC$_{71}$BM-rich | -0.798 ± 0.018 | 146 ± 16 | 0.867 ± 0.033 | 128 ± 33 |
| | M phase | -0.733 ± 0.010 | 111 ± 8 | 0.869 ± 0.014 | 142 ± 14 |

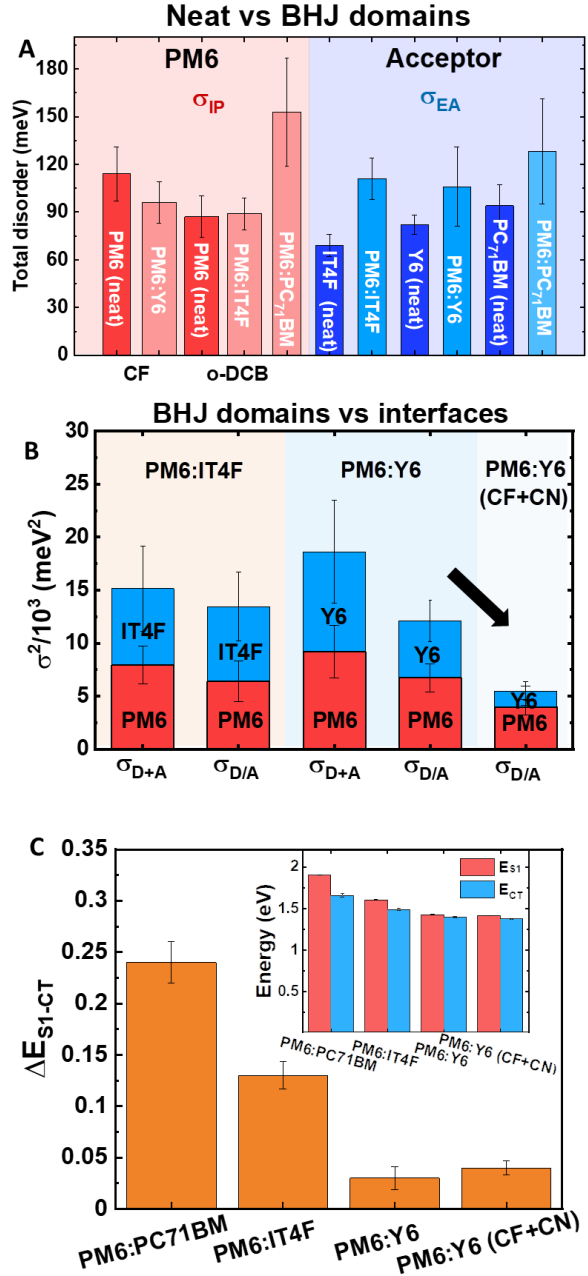

**Figure 3: Comparison of total disorder and $S_1$ to CT energy offsets**

(A) Comparison of total disorder obtained from the STM/S analysis of neat D and A materials and D-rich and A-rich phases in the BHJ. (B) Squared Gaussian width obtained from addition in quadrature of the disorder parameter from the nanoscale domains compared with the squared Gaussian width measured from the sharp interfaces. (C) Comparison of the singlet state energy ($E_{S1}$) and the CT state energy ($E_{CT}$) offsets ($\Delta E_{S1-CT}$) obtained using combined STM/S and s-EQE measurements, corresponding values are shown as inset, for the mentioned blend systems used in our study.

Comparison of $\sigma_T$ from the neat D and A materials to the D and A-rich phases in the BHJ for the studied cases (**Figure 3(A)**) suggests that the PM6 disorder is similar as it is mixed in the blend for PM6:Y6 and PM6:IT4F but increased for PM6:PC$_{71}$BM (compared with neat PM6 in oDCB) whereas the acceptor disorder increases in the blend for PM6:IT4F. Notably, most of the differences in the electronic disorder for various regions are within the 1-sigma error bars (Details of the uncertainty analysis are provided in **SI section 6**).

It is worth emphasizing that the interfacial disorder ($\sigma_{D/A}$) contributions of PM6 and IT4F to the squared Gaussian width ($\sigma_T^2/10^3$ (meV)$^2$) are similar to the disorder contributions of PM6 and IT4F calculated from domain interiors ($\sigma_{D+A}$) within the BHJ. One crucial observation in our study is that, in the case of PM6:Y6, the $\sigma_{D/A}$ is *lower* than $\sigma_{D+A}$ (the standard error are not overlapping), suggesting a reduced electronic disorder at the sharp interface compared to the interior of enriched domains (**Figure 3(B)**). This is a key observation that distinguishes PM6:Y6 from the other blends.

To further understand the influence of interfacial disorder on the overall performance, we varied the composition, formulation, and processing of the as-cast PM6:Y6 by using a chloronapthalene solvent additive (0.5% CN, wt. ratio (1:1.2)) followed by thermal annealing of the active layer. For simplicity, we refer to these samples as PM6:Y6(CF+CN). These efforts led to an increased fill factor from 60% to 74% and power conversion efficiency increase from 11% to 16%. The V$_{oc}$ increased from 0.8 V to 0.82 V, mostly due to the change in weight ratio (1:1.2) as compared to (1:1) in as-cast devices in the inverted structure used in our study. STM/S measurements at sharp interfaces in this blend (**Figure 3(B)**) reveal considerably reduced total interfacial energetic disorder from 110 ± 9 meV to 72 ± 7 meV **(Figure S21)**. STM/S analysis shows this reduction results from energetic disorder of both the PM6 IP and the Y6 EA declining at the PM6/Y6 interface. These results are consistent with the static disorder decline from **51 meV** to **44 meV** obtained from the apparent Urbach energy analysis (**Figure S22**).

We directly observe shifts in the IP and EA onsets in dI/dV spectra collected across the lateral sharp PM6/IT4F (**Figure S5**) and PM6/Y6 interfaces (**Figure S19**). This local band bending phenomenon is hypothesized to sweep charges away from the interface and facilitate the dissociation of CT states into free carriers[23,40]. These are particularly important observations in the case of the NFA's considered here due to their relatively low singlet to CT offset $\Delta E_{S_1\text{-}CT}$ which has caused significant issues in fitting the CT states, especially for PM6:Y6[45,58]. Therefore, obtaining the CS/CT state energy distribution independently using the combined STM/S and s-EQE helps to resolve the tricky issues of CT state energy estimation using s-EQE and EL measurements for such systems.

The hybridization of $S_1$ and CT states due to low $\Delta E_{S_1\text{-}CT}$ has been reported to lower $\Delta V_{nr}$[59]. However, it also results in a low driving force for CT state dissociation and can lead to incomplete exciton dissociation, reducing the charge generation efficiency[60]. So, a natural question is: How can the low $\Delta E_{S_1\text{-}CT}$ systems maintain a high EQE and simultaneously a low $\Delta V_{nr}$? In the case of modern NFAs, the electrostatic potential across a D/A interface, evidenced by the shifts in IP and EA onsets across the D/A interface (**Figure S19**), provides the needed driving force and may be controlled in the case of Y6 by the molecule's intrinsic quadrupole moment[23,61].

For the blend systems studied here, we report the $E_{S1}$ and $E_{CT}$ offsets and their values extracted from the combined STM and s-EQE methodology (**Figure 3(C)**). These measurements are consistent with those in the literature[45,48]. In addition, we directly show which interfaces are involved in charge generation and recombination.

**Table 2: STM/S total disorder analysis for PM6 based BHJ**

Peak position and standard deviation obtained from the statistically weighted Gaussian fits to the histograms of the different interfaces measured using STM/S (CS distribution) for the studied PM6 based BHJs

|  |  | Peak position (eV) | Total disorder (meV) |
|---|---|---|---|
| PM6:IT4F | $\Delta E_{D/A}$ | 1.483 ± 0.013 | 116 ± 14 |
|  | $\Delta E_{M/A}$ | 1.529 ± 0.016 | 131 ± 16 |
|  | $\Delta E_{D/M}$ | 1.605 ± 0.020 | 122 ± 19 |
|  | $\Delta E_{M/M}$ | 1.671 ± 0.016 | 139 ± 14 |
| PM6:Y6 | $\Delta E_{D/A}$ | 1.392 ± 0.010 | 110 ± 9 |
|  | $\Delta E_{M/A}$ | 1.436 ± 0.023 | 142 ± 26 |
|  | $\Delta E_{D/M}$ | 1.502 ± 0.025 | 124 ± 46 |
|  | $\Delta E_{M/M}$ | 1.598 ± 0.020 | 135 ± 21 |
| PM6:PC$_{71}$BM | $\Delta E_{D/A}$ | - | - |
|  | $\Delta E_{M/A}$ | 1.614 ± 0.028 | 177 ± 31 |
|  | $\Delta E_{D/M}$ | 1.637 ± 0.030 | 220 ± 30 |
|  | $\Delta E_{M/M}$ | 1.652 ± 0.020 | 162 ± 20 |

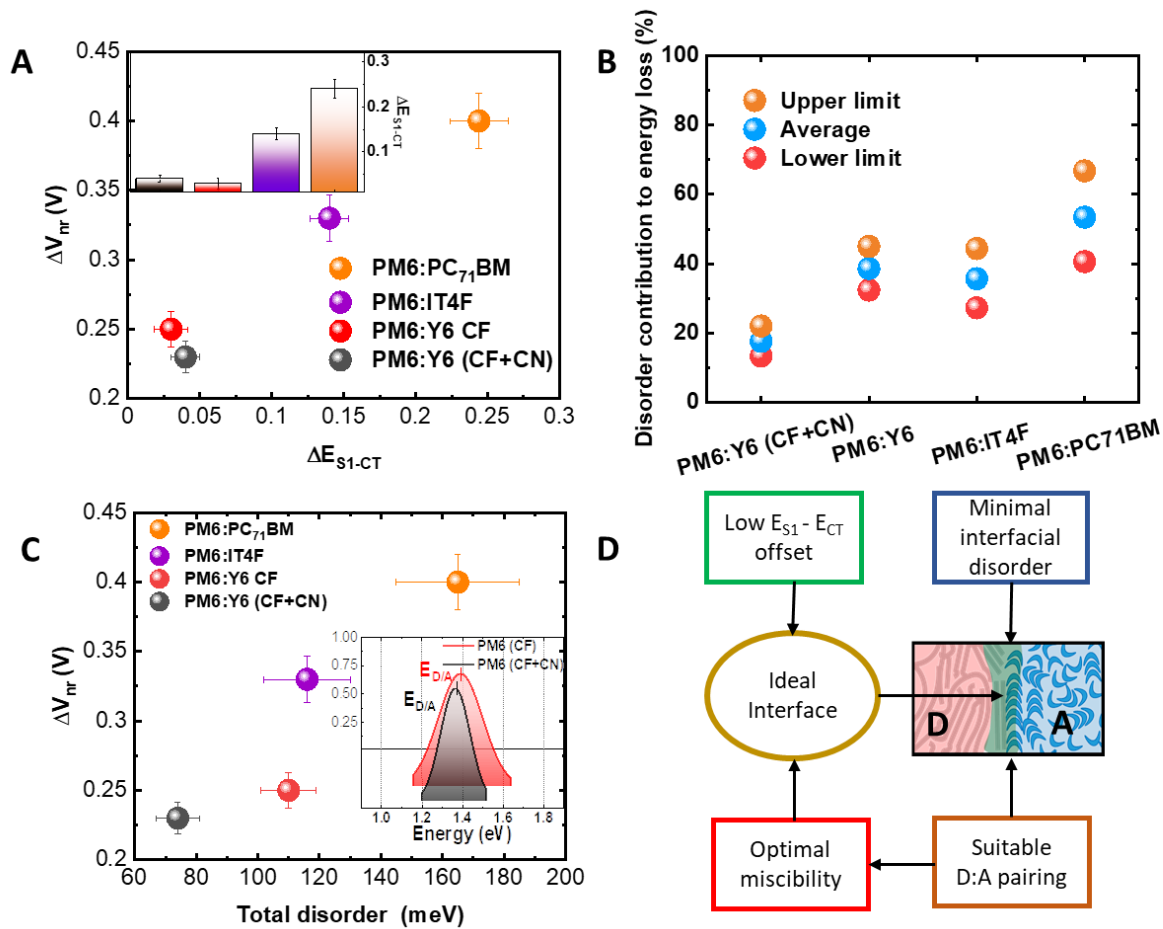

**Figure 4**: **Factors contributing to lower non-radiative voltage loss and disorder contribution to total energy loss**

(A) The non-radiative voltage loss (ΔV$_{nr}$) as a function of ΔE$_{S1-CT}$ obtained from the combined STM/S and s-EQE analysis for the different blend systems. (B) Contribution of the total disorder of the functional interface to the total energy loss obtained from the studied blend systems. (C) The non-radiative voltage loss (ΔV$_{nr}$) as a function of total disorder obtained from the combined STM/S and s-EQE analysis for the different blend systems studied here with inset highlighting the reduced total disorder compared between PM6:Y6 as-cast and optimized. (D) Schematic illustration of the factors leading to an ideal interface resulting in a low-loss OPV device.

## 2.c Correlation between Energetic Landscape and Nonradiative Voltage Losses

While both PM6:IT4F and PM6:Y6 are hypomiscible systems, the latter yields lower nonradiative voltage losses ($\Delta V_{nr} \cong 0.25$) than the former ($\Delta V_{nr} \cong 0.35$)[4,5]. Meanwhile, PM6:PC$_{71}$BM shows considerably higher nonradiative voltage losses ($\Delta V_{nr} \cong 0.4$) (see details of the voltage loss analysis in **SI section 8** and **Table S5**). The differences in non-radiative voltage losses for the three blend systems studied in this paper can be attributed to a combination of factors including the electronic disorder of the underlying CT states[42,45], the $\Delta E_{S1\text{-}CT}$ offset[5,59,60], and the kinetics of the recombination via triplet states[4]. The latter factor is shown to be essentially connected to the hybridization of charge transfer and singlet state governed by the $\Delta E_{S1\text{-}CT}$ offset[4].

We will first consider how non-radiative voltage losses are impacted in these systems with respect to $\Delta E_{S1\text{-}CT}$ and interfacial CT state disorder as measured using our framework. The $\Delta E_{S1\text{-}CT}$ for PM6:PC$_{71}$BM, PM6:IT4F, PM6:Y6 and PM6:Y6(CF+CN) studied here are approximately 0.24 eV, 0.13 eV, 0.03 eV and 0.04 eV respectively **(Figure S23)**, indicating a lowest S$_1$ and CT hybridization for PM6:PC$_{71}$BM followed by PM6:IT4F and highest for PM6:Y6. **Figure 4(A)** shows the $\Delta V_{nr}$ as a function of the $\Delta E_{S1\text{-}CT}$ for the blend systems studied here. We observe that reduction in $\Delta E_{S1\text{-}CT}$ leads to a decrease in $\Delta V_{nr}$. Interestingly, PM6:Y6 and PM6:Y6 (CF+CN) show similar $\Delta E_{S1\text{-}CT}$. However, a reduced $\Delta V_{nr}$ is observed for the latter case, which was previously shown to exhibit considerably lower interfacial energetic disorder. We further evaluate the influence of interfacial energetic disorder by varying the formulation and processing conditions of PM6:Y6, which allows the $\Delta E_{S1\text{-}CT}$ to be maintained constant and evaluating the impact of disorder alone on non-radiative voltage loss. We fabricated PM6:Y6 using oXY and oDCB in addition to the standard PM6:Y6 using CF and PM6:Y6(CF+CN) samples. We observe that PM6:Y6 shows variable non-radiative voltage loss (**Figure S24 and Table S7**) highlighting the role of CT state disorder upon the non-radiative voltage loss[8,42,45,48].

Interfacial CT state disorder can be used to calculate the contribution to the overall energy loss of the OSCs, estimated by the equation $E_{loss} = \sigma_T^2/2k_BT$[48]. Given the interfacial disorder values of the systems studied as shown in Table 2, the energy loss contribution can be estimated to be in the range of 0.08 – 0.13 eV, 0.2 – 0.27 eV, 0.2 – 0.32 eV and 0.39 – 0.64 eV for PM6:Y6 (CN), PM6:Y6, PM6:IT4F and PM6:P$_{71}$BM respectively. These values are approximately 13 – 22 %, 32 – 45 %, 27 – 44 % and 40 – 68 % for PM6:Y6 (CF+CN), PM6:Y6, PM6:IT4F and PM6:PC$_{71}$BM, respectively, as a share of the total voltage loss ($E_g^{PV}/q - V_{oc}$). We note that, in the case of the most disordered system, PM6:PC$_{71}$BM, 54±14% of the total energy loss can be attributed to energetic disorder through both radiative and non-radiative pathways, whereas in the highly optimized case of PM6:Y6(CF+CN), the share of energy loss due to energetic disorder is reduced to 17.5±4.5%. These observations show that reduction of interfacial disorder can increase the open-circuit voltage as summarized in **Figure 4(B)**. Extrapolation of $V_{oc}$ vs temperature data for the case of PM6:Y6[61] leads to a CT energy of 1.1 eV, which is still 0.3 eV higher than the $qV_{oc}$, showing that the disorder remains prevalent at room temperature even for some of the best performing systems like PM6:Y6.

In **Figure 4C**, we plot $\Delta V_{nr}$ against the total disorder of the functional interfaces as measured from the combined methodology **(Table S5)**. As expected, we observe that total disorder has a strong correlation with nonradiative voltage loss, in agreement with prior experimental work[42,45]. This observation stems from the fact that the rate of non-radiative recombination has exponential dependence on the total disorder as given by the equation[9],

$$k_{nr} = \frac{4\pi^2}{h}|V_{el}|^2 \frac{1}{\sqrt{2\sigma_T^2}} exp\left(\frac{(\lambda - E_{CT})^2}{2\sigma_T^2}\right) \quad (3)$$

where $V_{el}$ denotes the electronic coupling between the CT states and the ground state, $E_{CT}$ is the CT energy and $\sigma_T$ is the total disorder. The above equation underlines that both dynamic and static disorder can impact the nonradiative recombination rates and voltage losses since the above equation enters determination of $\Delta V_{nr}$ from established analysis of the electroluminescent $EQE_{EL}$ as shown below[9,12]:

$$\Delta V_{nr} = -\frac{K_B T}{q} \ln(EQE_{EL}) \quad (4)$$

$$EQE_{EL} = \frac{k_r}{p_e k_r + k_{nr}} \quad (5)$$

where $EQE_{EL}$ denotes the electroluminescence external quantum efficiency, $K_B$ denotes the Boltzmann constant; T, the temperature; q, the electron charge; and $p_e$, the probability for a photon generated by radiative recombination to escape from the device.

These results highlight the importance of controlling interfacial energetic properties, including the CT state disorder as a future pathway to achieve targeted reductions in nonradiative losses. Most importantly, our study has allowed CT states to be assigned to functional interfaces, namely the sharp D/A interfaces in PM6:IT4F and PM6:Y6, as well as to the M/M domains in the PM6:PC$_{71}$BM system. Overall, there are multiple factors that contribute to the ideal interface which leads to low loss solar cells. In addition to the minimal interfacial disorder ($\sigma_T$) as measured by STM/S, which our study has shown to contribute to the lower non-radiative voltage loss, we highlight that $\Delta E_{S1-CT}$ must be as low as possible, yet the interface needs to maintain a driving force for charge separation and an optimal miscibility arising from a suitable D:A pairing. All these factors work in conjunction to create an ideal interface (**Figure 4(D)**) which maximizes the performance while simultaneously minimizing the overall energy loss.

This study shows that $\Delta V_{nr}$ is impacted both by $\Delta E_{S1-CT}$ previously linked to the balance between rates of back charge transfer and charge dissociation[4] and by the interfacial energetic disorder of the functional interface and its associated CT states. In the future, combination of this framework with ultrafast spectroscopies will help our field ascertain exciton and charge dynamics to link them with the spatial, energetic, and state pictures of the nanoscale BHJ. In this regard, our methodological framework's ability to spatially map the BHJ, its energetic landscape and CT state manifold may become the cornerstone to providing functional 3D representations of BHJ morphologies with realistic spatial, energetic, and state information toward a high-fidelity digital twin for OPV devices[62].

### 3. Conclusions:

We have developed a framework capable of mapping the CT state manifold of BHJ solar cells quantitatively by combining local energetic landscape mapping via spectromicroscopy (STM/S) with global optoelectronic (s-EQE) measurements of CT states. In doing so, the framework (1) assigns CT states to nanoscale interfaces, (2) identifies functional interfaces of the BHJ solar cell, (3) measures full statistical distributions of CS/CT states for donor and acceptor components across each domain and interface, and (4) cross-validates static energetic disorders with the combined STM/S and s-EQE analyses with the apparent Urbach energy. Using this framework, we showed how avoiding hypermiscible material combinations, such as PM6 and $PC_{71}BM$, helps reduce the interfacial energetic disorder and nonradiative voltage losses. We also showed how hypomiscible D:A systems can help create sharp D/A interfaces. However, not all hypomiscible D:A pairings yield lower non-radiative voltage loss, as seen by comparing PM6:IT4F and PM6:Y6. The reduction of voltage losses in the PM6:Y6 blend is due to the combined effect of reduced $S_1$-CT offset and reduced interfacial disorder. We show how selectively varying the energetic disorder while fixing the $S_1$-CT offset establishes the link between voltage losses and interfacial energetic disorder. An ideal target D/A interface therefore emerges by virtue of optimal miscibility and specific molecular interactions, low S1-CT offset, and minimal interfacial disorder.

**Acknowledgements:**

This work was supported by the Office of Naval Research under award number N00014-20-1-2183. The authors would like to acknowledge the prior contributions of Dr. Sukumar Dey, Dr. Jingying Wang and Samanvitha Sridhar, for developing sample preparation and measurement protocols for plan-view STM/S measurements on 10-20 nm thick BHJ films. Device fabrication and related measurements (JV and s-EQE) were performed in Prof. Ade's Laboratory, a part of the ORaCEL cluster facility at NC State University.

**Author contributions:**

A.A. and D.B.D. conceived the scientific framework with the help of G.J.T and M.C.. G.J.T. designed experimental protocols, coordinated the experimental work, performed the STM/S measurements, and analyzed the STM/S data. M.C. fabricated the solar cell devices and performed the subsequent voltage losses and apparent Urbach energy analysis. G.J.T. and J.M. performed and analyzed the UV-vis data. G.J.T., M.C., A.A., and D.B.D. wrote the manuscript. All authors provided comments on the manuscript and contributed to the editing. A.A. directed the study.

**Data availability:**

All datasets generated and/or analyzed in this study are available from the corresponding author upon request.

56. Alam, S., Nádaždy, V., Váry, T., Friebe, C., Meitzner, R., Ahner, J., Anand, A., Karuthedath, S., De Castro, C.S.P., Göhler, C., et al. (2021). Uphill and downhill charge generation from charge transfer to charge separated states in organic solar cells. J. Mater. Chem. C *9*, 14463–14489. 10.1039/D1TC02351A.

57. Garcia-Belmonte, G., Boix, P.P., Bisquert, J., Lenes, M., Bolink, H.J., La Rosa, A., Filippone, S., and Martín, N. (2010). Influence of the Intermediate Density-of-States Occupancy on Open-Circuit Voltage of Bulk Heterojunction Solar Cells with Different Fullerene Acceptors. J. Phys. Chem. Lett. *1*, 2566–2571. 10.1021/jz100956d.

58. Shoaee, S., Luong, H.M., Song, J., Zou, Y., Nguyen, T., and Neher, D. (2023). What We have Learnt from PM6:Y6. Adv. Mater., 2302005. 10.1002/adma.202302005.

59. Eisner, F.D., Azzouzi, M., Fei, Z., Hou, X., Anthopoulos, T.D., Dennis, T.J.S., Heeney, M., and Nelson, J. (2019). Hybridization of Local Exciton and Charge-Transfer States Reduces Nonradiative Voltage Losses in Organic Solar Cells. J. Am. Chem. Soc. *141*, 6362–6374. 10.1021/jacs.9b01465.

60. Vezie, M.S., Azzouzi, M., Telford, A.M., Hopper, T.R., Sieval, A.B., Hummelen, J.C., Fallon, K., Bronstein, H., Kirchartz, T., Bakulin, A.A., et al. (2019). Impact of Marginal Exciton–Charge-Transfer State Offset on Charge Generation and Recombination in Polymer:Fullerene Solar Cells. ACS Energy Lett. *4*, 2096–2103. 10.1021/acsenergylett.9b01368.

61. Perdigón-Toro, L., Zhang, H., Markina, A., Yuan, J., Hosseini, S.M., Wolff, C.M., Zuo, G., Stolterfoht, M., Zou, Y., Gao, F., et al. (2020). Barrierless Free Charge Generation in the High-Performance PM6:Y6 Bulk Heterojunction Non-Fullerene Solar Cell. Adv. Mater. *32*, 1906763. 10.1002/adma.201906763.

62. Lüer, L., Peters, I.M., Smith, A.S., Dorschky, E., Eskofier, B.M., Liers, F., Franke, J., Sjarov, M., Brossog, M., Guldi, D.M., et al. (2024). A digital twin to overcome long-time challenges in photovoltaics. Joule *8*, 295–311. 10.1016/j.joule.2023.12.010.
35

**Supplementary information**

**Mapping Interfacial Energetic Landscape in Organic Solar Cells Reveals Pathways to Reducing Nonradiative Losses**

Gaurab J. Thapa, Mihirsinh Chauhan, Jacob P. Mauthe, Daniel B. Dougherty, Aram Amassian

**Supplementary Experimental Methods**

**Materials:**

All OPV materials were purchased from the commercial vendors: PBDBT-2F (PM6), BTP-4F (Y6) and (ITIC-4F) IT4F purchased form 1-Material, Zinc acetate dihydrate and Molybdenum (VI) oxide (MoO3) purchased from Sigma Aldrich, $PC_{71}BM$ was purchased from Solarmer Materials Inc. All the materials and solvents were used as received without further purification.

**Device preparation:**

All bulk heterojunction (BHJ) devices studied were fabricated with an inverted structure of indium tin oxide (ITO)/ZnO/polymer: NFA/MoO3/Al. Patterned ITO-glass were cleaned by successive sonication in soap solution (20 min), deionized water (15 min), acetone (15 min), and isopropanol (15 min). The cleaned substrates are treated with UV-ozone for 10 min and ZnO electron transport layer (ETL) was spin coted at 6000 rpm from precursor solution (Zinc acetate dihydrate was dissolved in anhydrous 2-Methoxyethanol and Ethanolamine) and annealed at 180 °C for 30 min. All BHJ active layers with standard thickness (~100 nm) were spin coated from at least 4h stirred solutions in a nitrogen-filled glovebox to obtain desired thickness confirmed by profilometry measurements. Formulation of Donor:Acceptor blend solutions used for casting of the active layer were 20mg/ml in o-DCB for PM6:IT4F and PM6:$PC_{71}BM$ (1:1 wt. ratio). For PM6:Y6, formulation in CF is prepared at 16mg/ml and 1:1 wt. ratio, while o-DCB and o-XY formulations are prepared at 20mg/ml and 1:1 wt. ratio. For the optimized PM6:Y6(CF+CN), formulation is prepared in CF + 0.5% v/v CN at 16mg/ml and 1:1.2 wt. ratio, followed by annealing at 100 °C for 5 min. For all blends, thin BHJ layers were fabricated using oDCB as the solvent unless otherwise stated with similar active layer thickness (15-20nm) used for STM/S study as confirmed by profilometry measurements, by using the same stock solution to prepare the standard thickness devices by increasing the spin rate (Table S1) to obtain thin active layers. Finally, 8 nm MoO3 and 100 nm Ag were evaporated under vacuum conditions of $1 \times 10^{-6}$ Torr to finish the device fabrication process, resulting in a device area of 7.6 mm$^2$ with evaporation shadow mask (Dimension of mask = 1.52 × 5.0 mm$^2$).

**Table S1: Spin coating parameters for active layers used in device preparation**

| Blends | Concentration | Solvent | Wt. ratio | Spin speed |
|---|---|---|---|---|
| PM6:IT4F (100 nm) | 20mg/ml | oDCB | 1:1 | 1500 rpm |
| PM6:IT4F (20 nm) | 20mg/ml | oDCB | 1:1 | 6000 rpm |
| PM6:PC$_{71}$BM (100 nm) | 20mg/ml | oDCB | 1:1 | 1500 rpm |
| PM6:PC$_{71}$BM (20 nm) | 20mg/ml | oDCB | 1:1 | 6000 rpm |
| PM6:Y6 (100 nm) | 20mg/ml | oDCB | 1:1 | 2000 rpm |
| PM6:Y6 (20 nm) | 20mg/ml | oDCB | 1:1 | 8000 rpm |
| PM6:Y6 (100 nm) | 20mg/ml | oXY | 1:1 | 2200 rpm |
| PM6:Y6 (100 nm) | 16mg/ml | CF | 1:1 | 2500 rpm |
| PM6:Y6 (20 nm) | 16mg/ml | CF | 1:1 | 9000 rpm |
| PM6:Y6 (100 nm) | 16mg/ml | CF + 0.5% CN | 1:1.2 | 2250 rpm |

**Device Characterization:**

For device characterizations, J–V curves were measured using a Keithley 236 source meter under AM1.5G light (100mWcm$^{-2}$) and using a Class AAA Newport solar simulator in a nitrogen filled glovebox, where light intensity was calibrated using a standard reference Si diode with KG5 filter purchased from PV Measurement. All J-V curves measured with forward scanning from −2.0 V to +1.0 V with a voltage step of 0.02V and a scan speed of 50mVs$^{-1}$ was used. J–V characteristics were recorded using a Keithley 236 source meter unit. The s-EQE spectra of the OPVs are measured using FTPS PECT-600 (Enlitech). The photocurrent response of the OPV was amplified and modulated using a lock-in-amplifier, where incoming photon flux incident is estimated using calibrated Si and Ge photodetectors.

**Atomic Force Microscopy (AFM):**

Atomic force microscopy measurements were acquired by using a commercial instrument (Asylum MFP-3D) and tips (Budget Sensors TAP300 E-G) in non-contact tapping mode.

All AFM images were acquired using a 1 Hz scan rate with 512 pixels per line. The AFM data is analyzed using Gwyddion.

1. Details of sample preparation and measurements using STM/S

The film processing condition parameters for the neat and blend materials are shown in Table S2. The spin-coating was performed on a glove box environment for 30 seconds. After spin-coating, the films were dried under vacuum for 30 minutes to let excess solvent evaporate from the film. Then, the spin-coated films were transferred to the load-lock chamber ($1 \times 10^{-6}$ Torr) of scanning tunneling microscopy/spectroscopy (STM/S) unit and stored for 24 hours for outgassing any remaining solvent and removal of any contaminations from the surface followed by transfer to the STM/S chamber ($5 \times 10^{-10}$ Torr). STM/S measurements were carried out inside a home-built UHV system using a commercial STM (Omicron VT-XA100/500).

**Table S2: Film preparation condition**

| Material | Total Concentration | Solvent | Weight ratio | Spin speed |
|---|---|---|---|---|
| PM6 | 4 mg/ml | oDCB | | 6000 rpm |
| Y6 | 8 mg/ml | oDCB | | 4000 rpm |
| PM6 | 2 mg/ml | CF | | 4000 rpm |
| Y6 | 1mg/ml | CF | | 3500 rpm |
| IT4F | 12 mg/ml | oDCB | | 1500 rpm |
| $PC_{71}BM$ | 8 mg/ml | oDCB | | 3000 rpm |
| PM6:Y6 | 4 mg/ml | oDCB | 1:1 | 4000 rpm |
| PM6:Y6 | 2 mg/ml | CF | 1:1 | 3000 rpm |
| PM6:Y6 | 2 mg/ml | CF + 0.5% CN | 1:1.2 | 2500 rpm |
| PM6:IT4F | 8 mg/ml | oDCB | 1:1 | 3000 rpm |
| PM6:$PC_{71}BM$ | 8 mg/ml | oDCB | 1:1 | 4000 rpm |

In the constant current mode of a STM measurement, a very sharp tip made from Pt/Ir (80/20), using mechanical method, is brought close to the sample, typically on the order of angstroms, and a bias voltage is applied between the tip and the sample. This bias voltage raises the Fermi energy of the sample/tip (depending on the bias voltage) such that the electrons tunnel between the tip and the sample. This tunneling current is monitored and maintained at a fixed value through a feedback loop. As the tip raster over the sample's surface, the surface features cause variation in tunneling current. To maintain a constant current, the distance between the tip and sample is varied, which is then used to construct the topography of the surface. For a single STS measurement, the tip is held over a fixed spatial location over the sample and the sample bias is swept from – 2V to 2V while measuring the tunneling current. For a semiconductor, the I-V curve consists of regions of suppressed current that corresponds to the electronic gap followed by ohmic region. Figure S1(A) shows an I-V curve measured from a neat PM6 sample on Au (111) substrate. This tunneling current can be expressed as[1–3],

$$I = \frac{4\pi e}{\hbar} \int_0^{eV} T(d,V,\epsilon) \rho_s(E_F + \epsilon) \rho_t(E_F - eV + \epsilon) \, d\epsilon \qquad (1)$$

where $T(d,V,\epsilon)$ is the tunneling transmission coefficient, $\rho_s$ and $\rho_t$ are the sample and tip density of states, respectively, $E_F$ is the Fermi energy, e and ℏ are elementary charge and reduced Planck's constant, respectively. The tunneling current is measured with reference to the Fermi level. With the tip density of state held constant, we obtain,

$$\frac{dI}{dV} \propto T(d,V,\epsilon) \rho_s(eV) \qquad (2)$$

Differential conductance (dI/dV) curves were obtained by using a lock-in amplifier. The dI/dV curve obtained for the neat PM6 is shown in Figure S1(B). The dI/dV curve highlights the onset region from which the tunneling current enters the linear regime. Using linear extrapolation, we can accurately measure the occupied and unoccupied frontier orbitals levels at the point of measurement, also referred to as the ionization potential (IP) and electron affinity (EA) levels respectively.

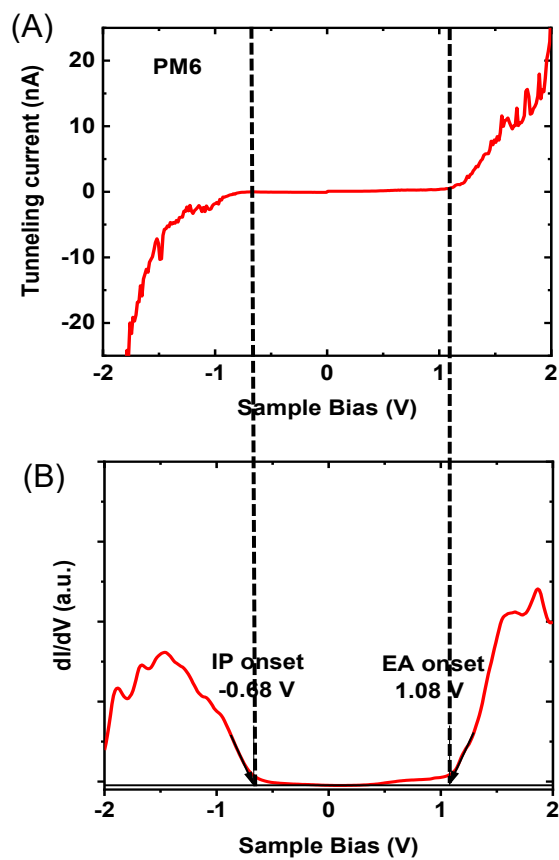

**Figure S1**: (a) A typical I-V spectra of tunneling current measured as a function of bias voltage applied to the neat PM6 (CF) sample. (b) The dI/dV spectrum of the corresponding I-V curve is obtained using a lock-in amplifier (Stanford Research) by applying a modulation voltage of 100mV, 3 ms time constant and 10 KHz modulation frequency. The IP and EA onset are obtained by extrapolating the linear regime of the dI/dV spectrum.

## 2. dI/dV spectra of neat films and comparison with UPS/IPES

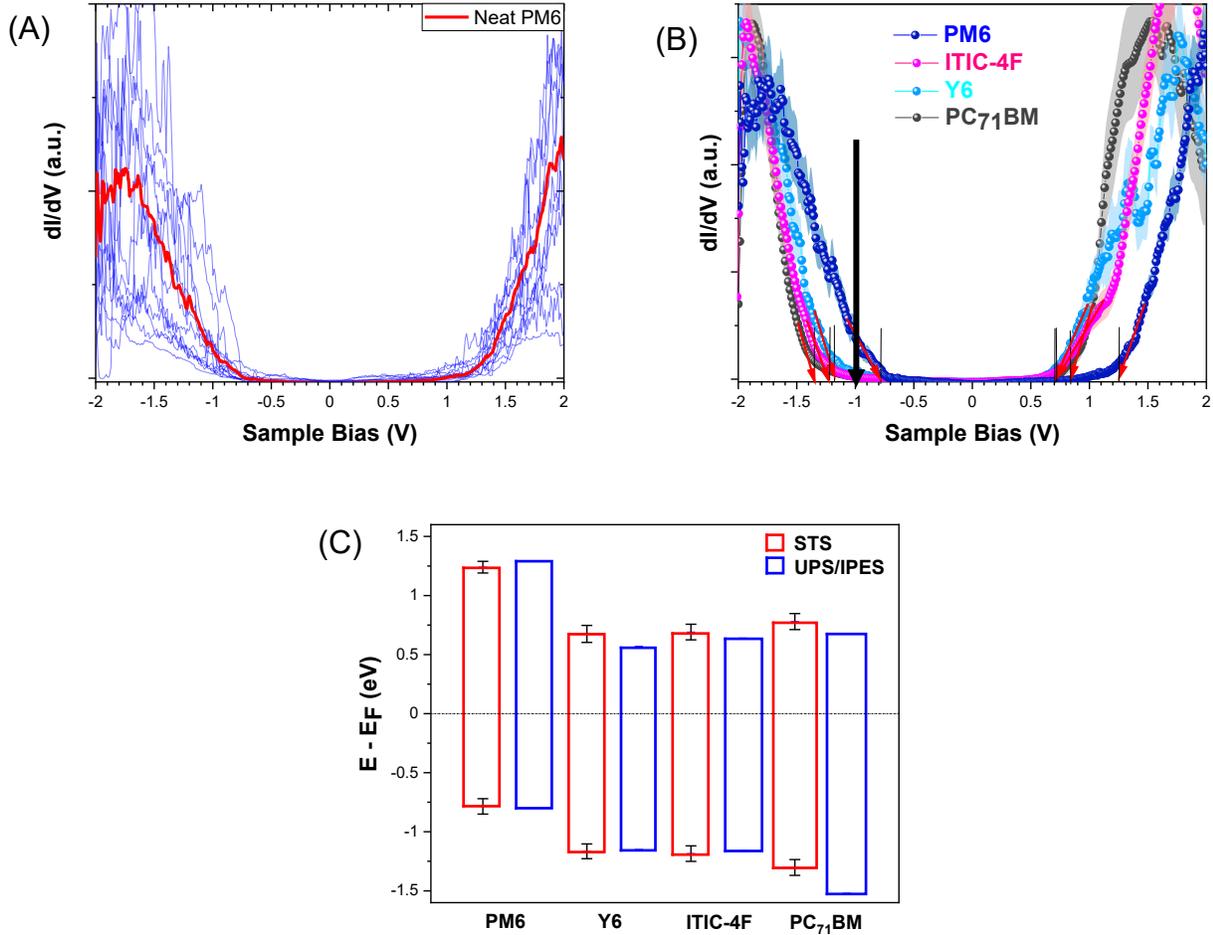

**Figure S2**: (A) Process of averaging dI/dV spectra from different location to get averaged dI/dV spectrum shown for neat PM6 where the averaged dI/dV spectrum is shown by red line and the dI/dV spectra used for averaging are shown by blue lines. (B) Averaged dI/dV spectrum of neat PM6, Y6, IT4F and PC$_{71}$BM. The IP and EA onset are obtained by extrapolating the linear regime of the averaged dI/dV spectrum. (C) Comparison of IP and EA onsets from STM/S with UPS/IPES from literature[4].

STM/S is a localized form of measurement and sensitive to the point of measurement. Therefore, it is essential to perform the measurements at different spatial locations. Around 12-15 dI/dV spectra are collected from various spatial locations in neat films.

Then, they are averaged to get a single dI/dV spectrum as shown in Figure S2(A). By comparing the averaged dI/dV spectra of the neat materials along with the standard error shown in the shaded region as shown in Figure S2(B), we can identify a suitable bias voltage to probe the BHJ. Here, we choose a suitable bias voltage of -1V to probe our BHJ so that we can obtain a reasonable contrast in conductance associated with the donor(D) and acceptor(A)-rich phase in the BHJ. Figure S2(C) shows comparison of IP and EA onsets from STM/S measurements of neat films with UPS/IPES data from literature[4].

## 3. Identifying the D-rich, A-rich, M phase and different interfaces using STM/S

Spectroscopically, the D (A)-rich phase exhibits a p-type (n-type) characteristic with an IP (EA) level closer to the Fermi level. The spectra from the M phase shows both donor IP feature and acceptor EA feature with some offset due to changes in the aggregation state and local environment of molecules[5,6]. Figure S3(A) shows a dI/dV map of PM6:IT4F acquired at -1V which is sensitive to the IP region on the donor PM6. Therefore, the area of high conductance corresponds to the occupied region of PM6 and similarly the area of low conductance corresponds to the unoccupied region of IT4F and therefore can be associated with the respective phase in the blend.

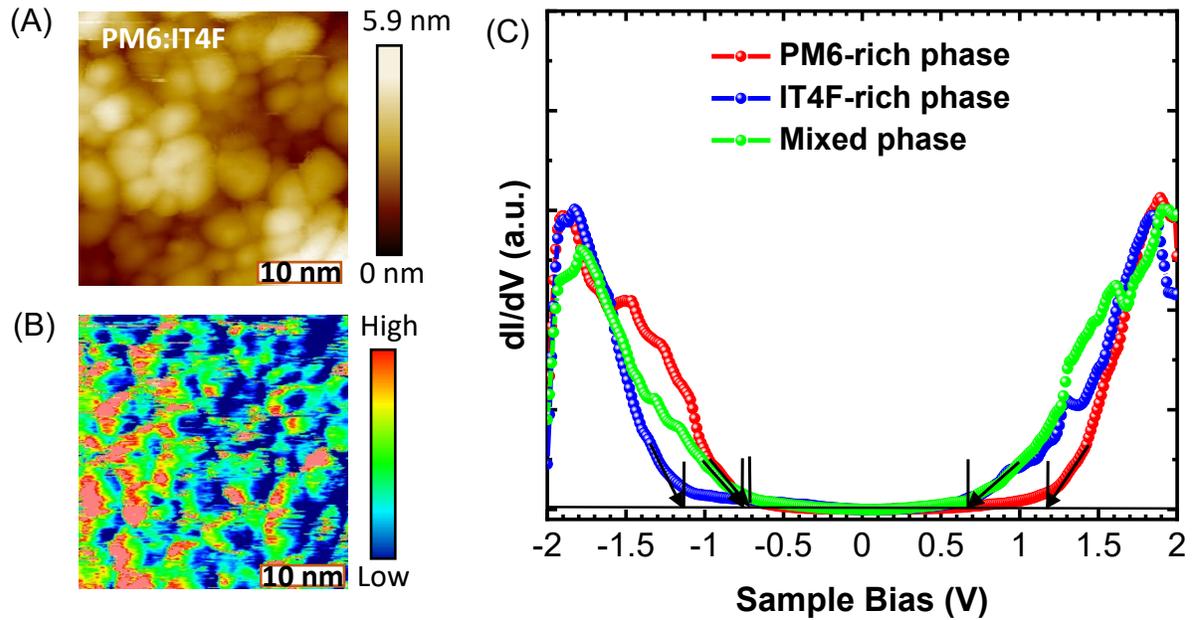

**Figure S3**: (A) Topography of PM6:IT4F measured under (-1V, 0.5 nA) tunneling setpoints. (B) dI/dV maps associated with the topographic map of PM6:IT4F showing areas of higher, lower, and intermediate conductance. (C) Averaged dI/dV spectra collected from the red, blue, and green areas of various dI/dV maps corresponding to the PM6, IT4F-rich phase and the M phase.

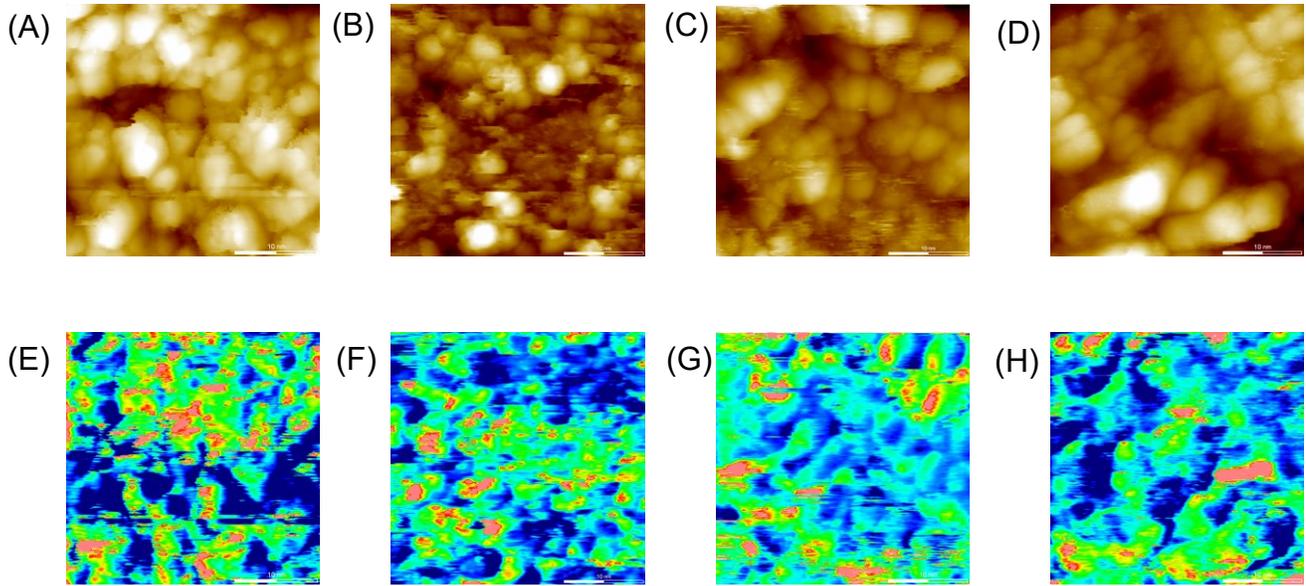

**Figure S4:** Selected topographical map (A-D) and corresponding dI/dV maps (E-H) of PM6:IT4F used to collect the dI/dV spectra from the PM6-, IT4F-rich and M phase.

For acquiring dI/dV spectra along a line, a suitable area is selected based on the measurements. Repeated scans are performed to ensure that no tip artifacts are present that might hamper the measurements. Once it is ensured that the tip is stabilized, an area of contrast is chosen according to the contrast in the dI/dV maps. Then, the line scans are performed from the area of high contrast to low contrast or vice versa. Those dI/dV spectra are compared to the dI/dV spectra from the D, A-rich phases and thus the domains representing the D, A-rich phases is identified from the line scans. In between the D,A-rich phase, represented by the narrow green area, the sharp interface exists. Based on the dI/dV spectral changes (offset in the EA and IP onsets from the D and A rich phase respectively) compared to the D,A-rich phases, dI/dV spectra of the interface areas are chosen to construct a histogram.

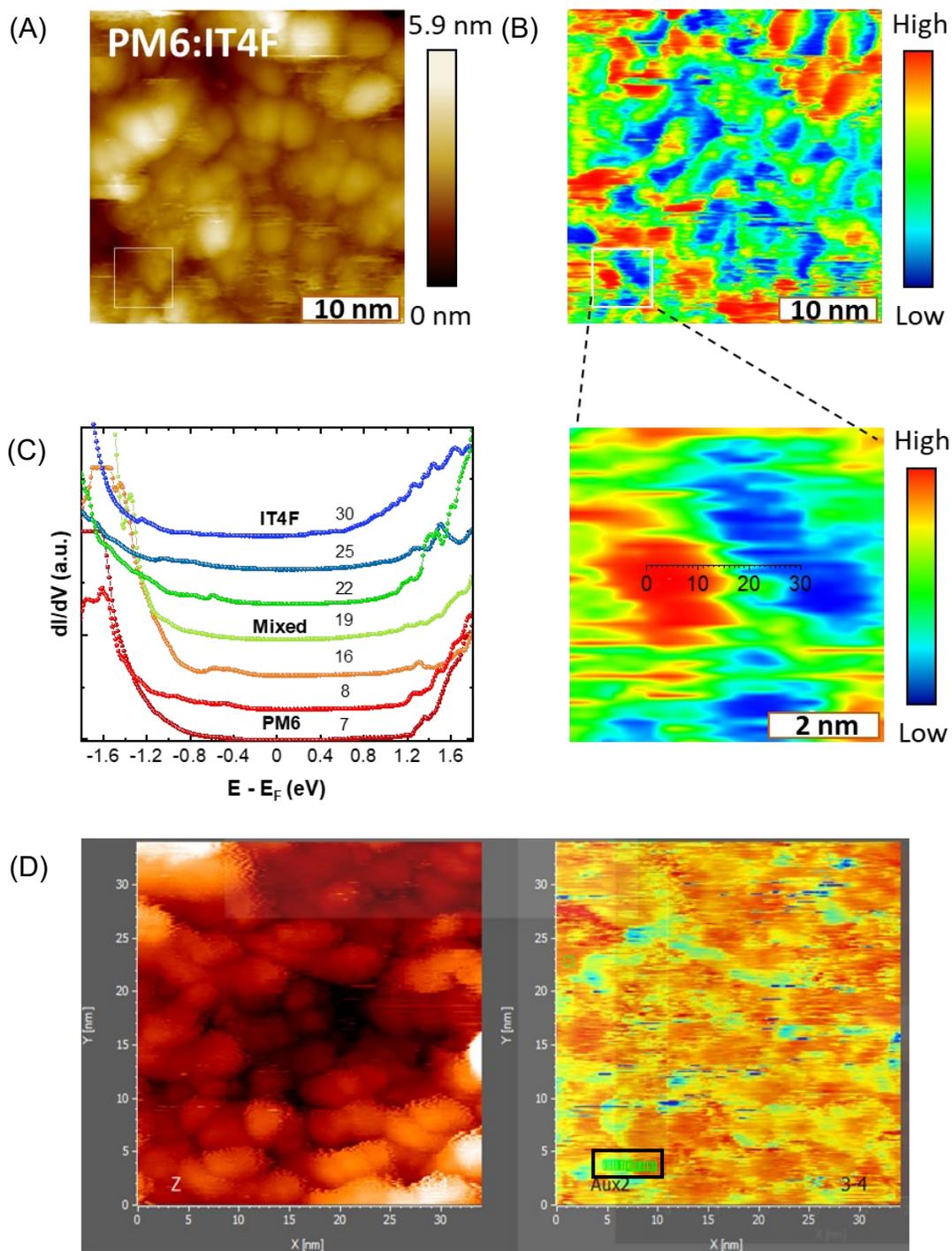

**Figure S5**: (A) Topography and (B) dI/dV map of PM6:IT4F and a magnified section of the dI/dV map where spectroscopic measurement is performed along a line probed at (-1V, 500 pA) tunneling setpoints. The region covered in red, and blue are energetically identified as PM6 and IT4F-rich phase respectively with green region associated with the sharp interface with energetic offset as compared to the PM6 and IT4F spectra. (C) The

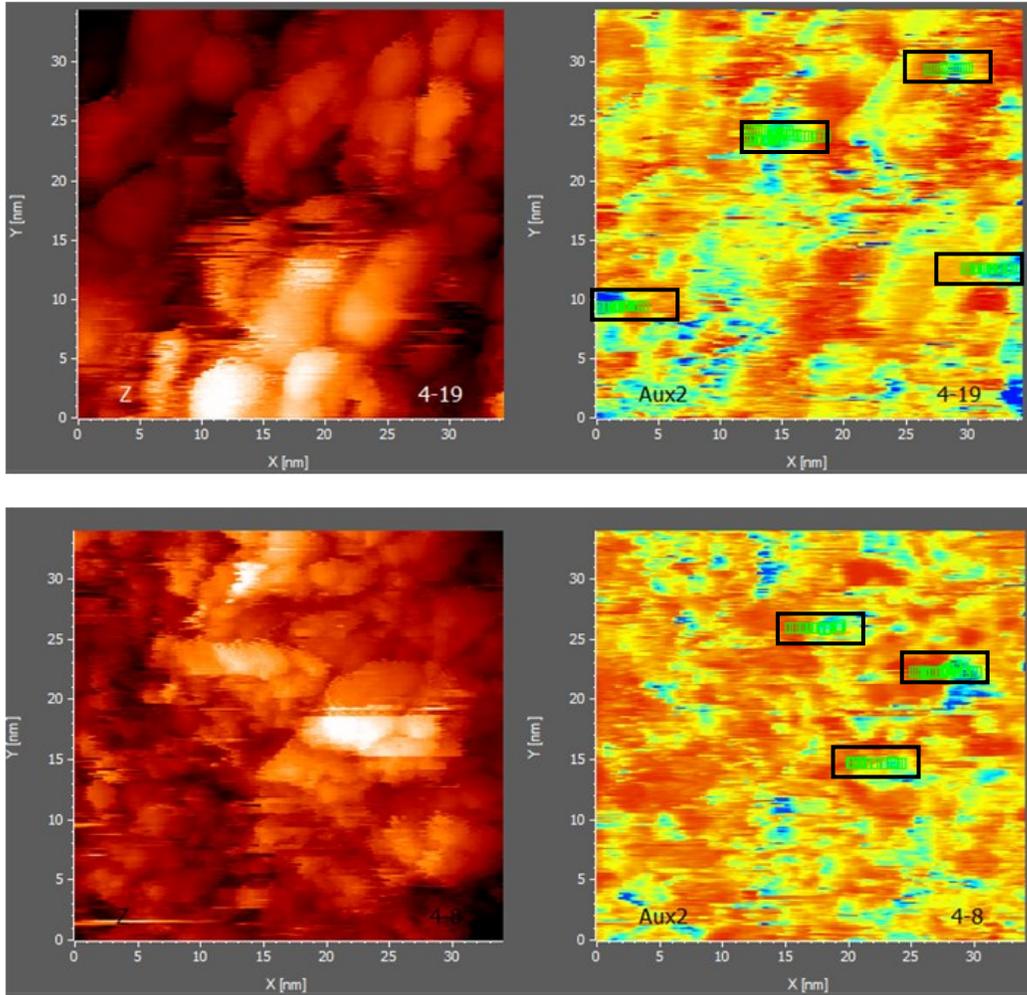

dI/dV spectra obtained from the line spectroscopy measurement. (D) The unprocessed topography and corresponding dI/dV maps from selected regions in PM6:IT4F for dI/dV spectroscopy across different interfaces. The points selected for the line scans are overlaid on the dI/dV maps.

## 4. Weighted fits to the histograms obtained from PM6:IT4F

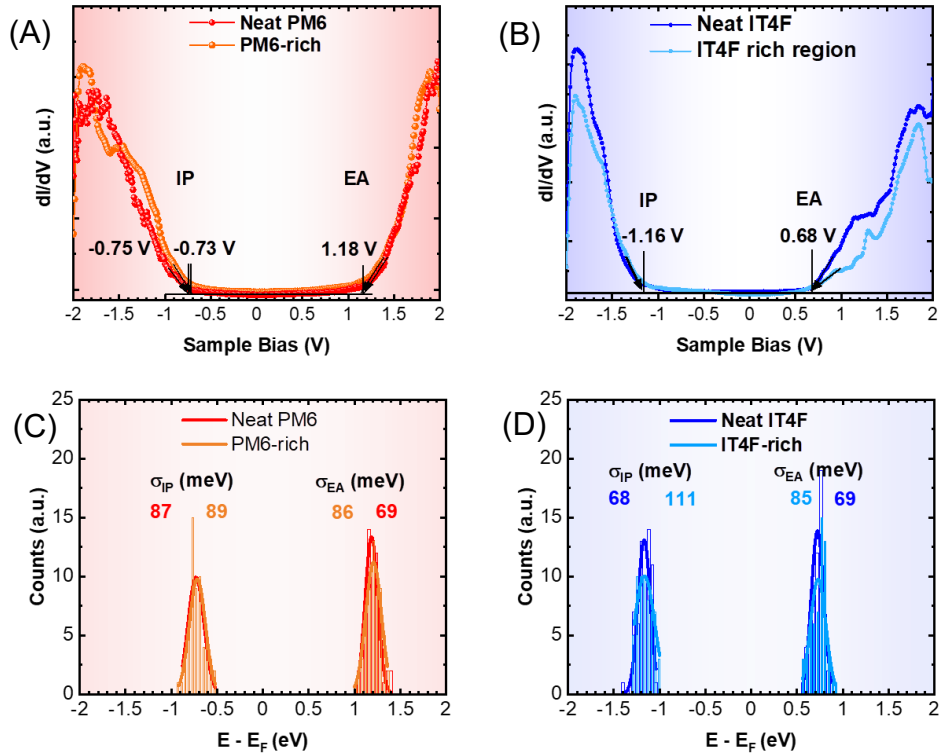

**Figure S6**: (A) & (B) Averaged dI/dV spectra of neat PM6 and IT4F as compared with the PM6 and IT4F-rich phases in the PM6:IT4F blend respectively. (C) & (D) Histogram obtained from the IP and EA onsets of neat PM6 and IT4F compared with the PM6 and IT4F-rich phases in PM6:IT4F blend respectively. The histograms were compiled from 50-60 dI/dV spectra acquired from different spatial positions in the neat and blend film corresponding to the PM6 and IT4F-rich phases in the PM6:IT4F blend using 0.04 eV bins.

The dI/dV spectra acquired at various spatial location within a D, A rich and M phase can vary in terms of their energy onsets depending on the local conformation and interaction with neighboring molecules. This spectral variation can provide an estimation of the local electronic disorder of the D, A, and M phase within a BHJ. Statistical analysis of these measurements enables us to create a histogram of band edges as shown in Figure S6 that can provide insight into the electronic disorder associated with these different phases.

Since the dI/dV spectra are proportional to the local density of states (LDOS) of the material as shown in equation 2, these histogram of band edges are a proxy measurement for the LDOS and associated electronic disorder.

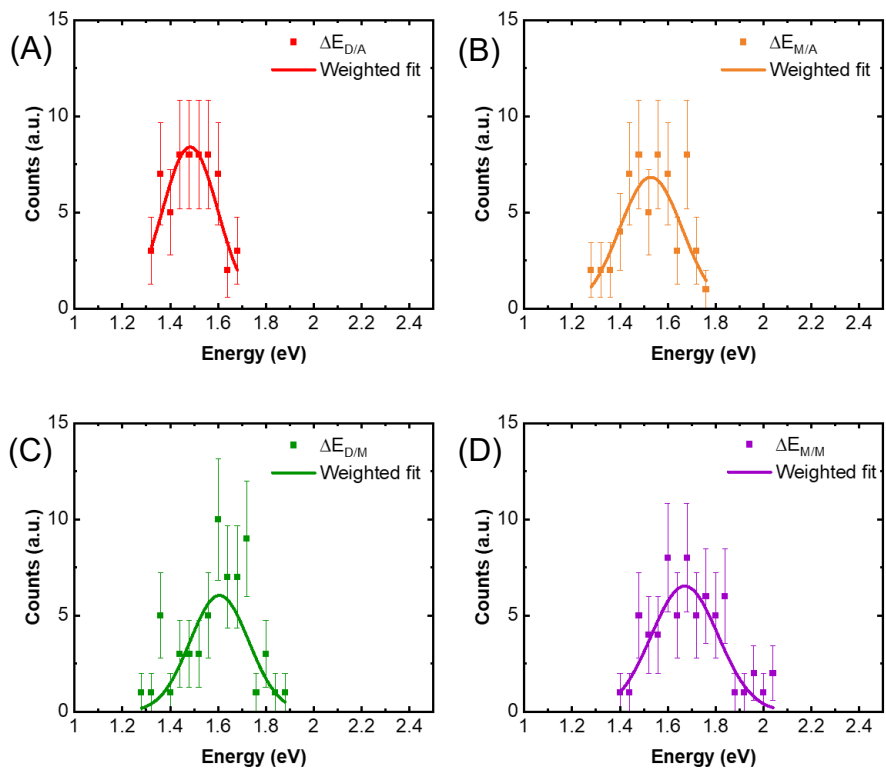

**Figure S7**: Weighted Gaussian fits to the histogram obtained from the energetic difference of the IP and EA from the (A) sharp interfacial region, (B) M phase and IT4F-rich phase, (C) PM6 and M phase and (D) M phase respectively. The data points are the number of occurrences using 0.04 eV energy bins with Poisson error bars.

## 5. Determination of CT state properties & static disorder

**Approach 1: Static disorder using combine s-EQE and STM/S approach.**

We measure s-EQE using FTPS PECT-600 as shown in Fig. S15 for blend. We also measured neat material s-EQE to make sure that we are fitting the correct regime of EQE tail of blends. We used modified Marcus theory[7–9] (Equation 3) to fit the tail of s-EQE to determine CT state properties by incorporating distribution energy as a $E_i$ (i: D/A, M/A, D/M and M/M) energy and total disorder $(\sigma_T^i)$ from gaussian fit to the histogram of the energetic difference between IP and EA onsets of the associated phases measured using STM/S as shown in Fig S9 in the case of PM6:IT4F.

$$EQE = \frac{f_i}{E\sqrt{2\pi(\sigma_T^i)^2}} exp\left(-\frac{(E_i + \lambda_i - E)^2}{2(\sigma_T^i)^2}\right) \quad (3)$$

where $\lambda_i$ is the reorganization energy, $f_i$ is the oscillator strength of the i[th] CT state, and E is energy. The total disorder $(\sigma_T^i)$ is the sum of the static $(\sigma_s^i)$ and dynamic disorder $(\sigma_d^i)$ and related to the reorganization energy as,

$$(\sigma_T^i)^2 = (\sigma_s^i)^2 + (\sigma_d^i)^2; \ (\sigma_d^i)^2 = 2\lambda_i K_B T \quad (4)$$

where $K_B$ and T are the Boltzmann constant and temperature respectively. Except for the oscillator strength, all of the parameters in this expression can be accurately estimated from the CS state proxies obtained in our STS measurements. Comparing experimental EQE measurements, we can associate the CT states manifold to the various interfaces visualized by STM/S and directly identify which of the interfaces host the CT states that are most relevant to recombination under open circuit voltage conditions.

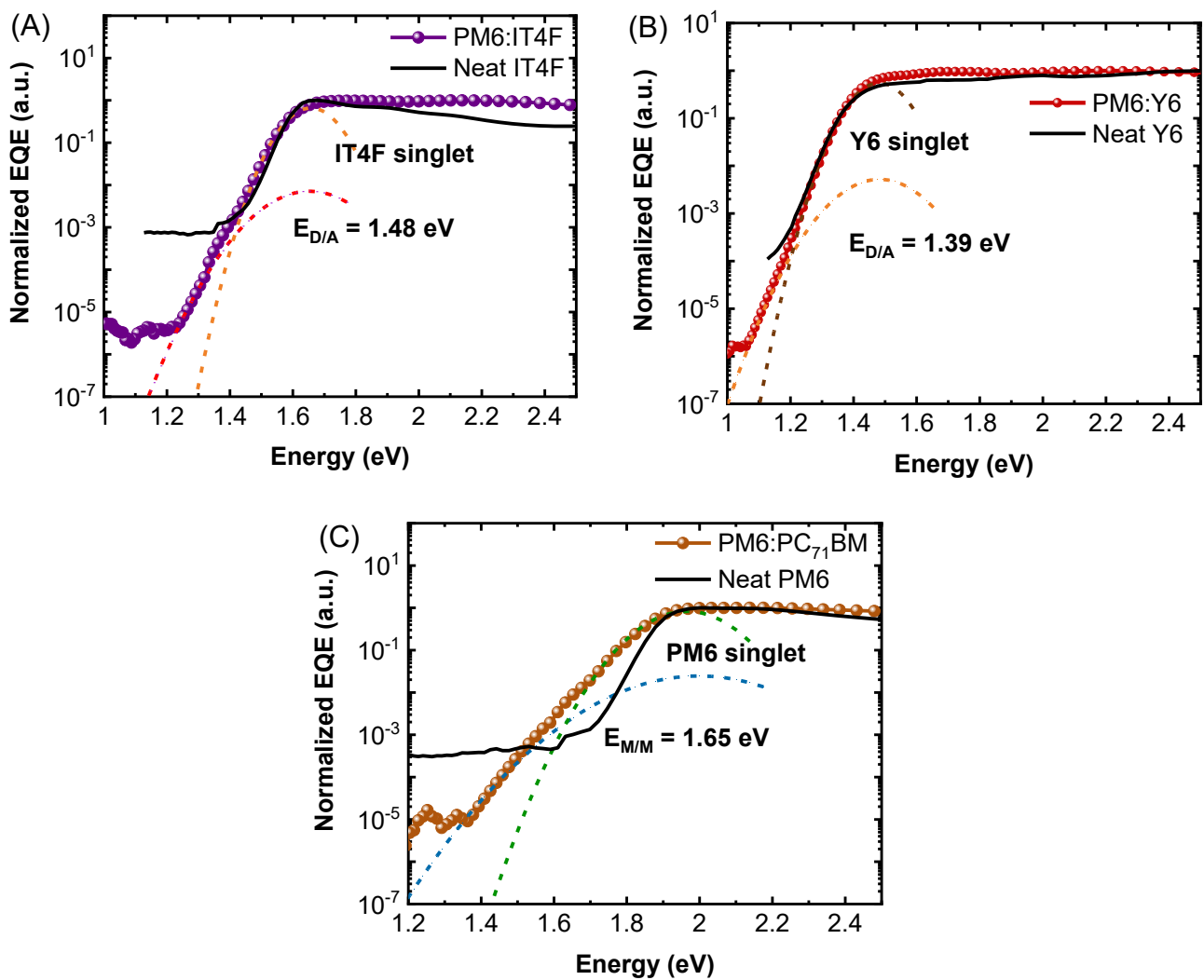

**Figure S8:** S-EQE of neat PM6, IT4F and Y6 were plotted for the comparison with blend s-EQE of (A) PM6:IT4F, (B) PM6:Y6 and (C) PM6:PC71BM respectively. The tail of sEQE were fitted using modified Marcus theory (explained in **Approach 1**) to determine CT state properties by incorporating distribution energy as a $E_{D/A}$ energy and total disorder ($\sigma_T$) from weighted gaussian fit to the histogram of the energetic difference between IP and EA onsets of the associated phase measured using STM/S. Here, we notice a steeper absorption profile for the PM6:Y6 followed by PM6:IT4F whereas a broader absorption profile for PM6:PC71BM **(Figure S9)**. This is in part due to the S1-CT hybridization causing CT states to overlap with the singlet absorption.

**Approach 2: Static disorder using apparent Urbach energy from s-EQE**

We evaluate the static disorder value derived from apparent Urbach energy as described by Kaiser et. Al[9]. The apparent Urbach Energy is given as:

$$E_U^{app} = \left(\frac{d\ln(EQE)}{dE}\right)^{-1} \quad (5)$$

Near the bandgap, $E_U^{app}$ can be modelled as:

$$E_U^{app} = \frac{2\sigma_s^2}{E_X - E} \quad (6)$$

where $\sigma_s$ is the static disorder and $E_X$ is a fitting constant. From the apparent Urbach Energy, we observed that for PM6:PC$_{71}$BM static disorder derived near the optical bandgap of PM6 indicating that s-EQE is dominated by PM6 suggesting a mixed morphology of PM6:PC$_{71}$BM consistent with the STM/S analysis as described in main text. On the other hand, in the case of PM6:IT4F and PM6:Y6, s-EQE is dominated by low energy CT$_1$ state lies near the S$_1$ state of IT-4F and Y6, respectively, in agreement with the STM/S analysis.

**Table S3:** The determined CT state parameters and disorder values using the s-EQE and STM/S measurements:

| Active Materials | E$_{CT}$ (eV) | λ (eV) | σ$_t$ (meV) | σ$_s$ (meV) E$_U$ method | σ$_s$ (meV) s-EQE & STM/s |
|---|---|---|---|---|---|
| PM6:IT4F (o-DCB) | 1.483 ± 0.013 | 0.195 | 116 ± 14 | 51 ± 1.8 | 58 ± 14 |
| PM6:Y6 (CF) | 1.392 ± 0.010 | 0.175 | 110 ± 9 | 48 ± 0.5 | 55 ± 9 |
| PM6:PC$_{71}$BM (o-DCB) | 1.652 ± 0.020 | 0.36 | 162 ± 20 | 65 ± 2.2 | 87 ± 20 |

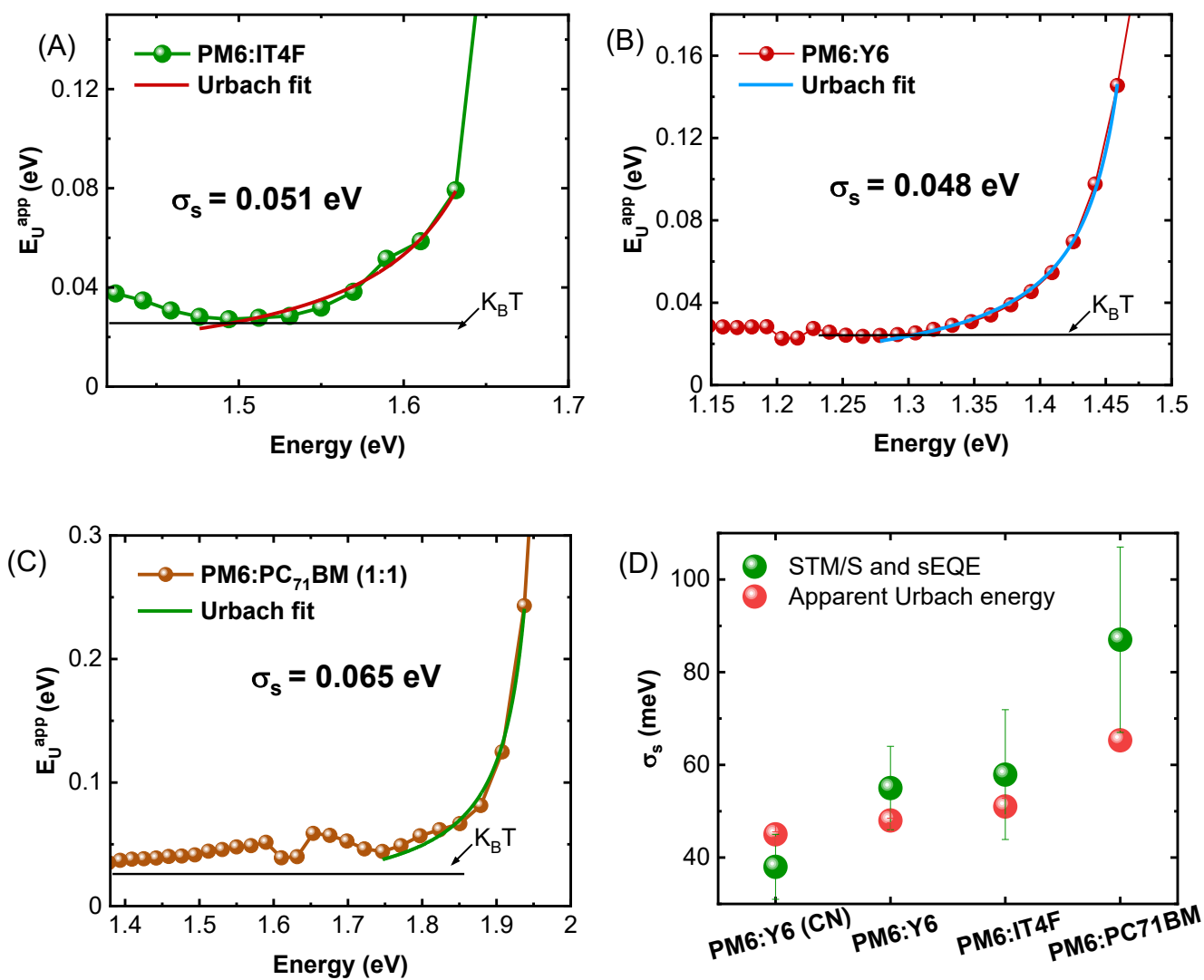

**Figure S9:** Apparent Urbach energy as a function of energy for (A) PM6:IT4F, (B) PM6:Y6 and (C) PM6:PC$_{71}$BM. Urbach fit was performed as explained for the blends studied (see **Approach 2**). (D) Comparison of static disorder obtained from the combined STM/S and sEQE analysis and apparent Urbach energy analysis.

## 6. Uncertainty propagation in STM/S measurements

The uncertainty values shown in figures 2 (D), (H), and (L) and Figure 3 (A) and (B) of the main manuscript are calculated considering the disordered nature of these organic semiconductors. Here, we will discuss how these uncertainty values are obtained using statistically weighted Gaussian fit to the histogram of the IP and EA offsets measured using the STS measurement on the different phases and interfaces.

The error bars are due to the nature of the experiment such as the localized form of measurement at room temperature. It should be noteworthy that as shown in Fig. 1 C, the energetic distribution used for the difference in the D/A interface and within the M domain are obtained from line spectroscopy across the D/A interface and local spectroscopy within the M domain, respectively. Similarly, the M/A and D/M interfaces are obtained from the difference between the M phase and A-rich spectra and the D-rich spectra. These are different spectra as compared to the D/A interfaces. They are spectra obtained from multiple STM maps of the A-rich and D-rich domains and M domains within the same BHJ.

The standard error of a linear fit to extrapolate the energetic offset LUMO offset ($E_F - E_C$) or HOMO offset ($E_V - E_F$) using the dI/dV spectra ranges from 0.01 – 0.02 eV. Considering the accuracy of the STS measurements, the width of the histograms is adjusted to 0.04 eV. This takes care of the uncertainty associated with the systematic error of the measurement. Regarding the uncertainty in the number of counts, it is governed by Poisson distribution where the square root of the number of occurrences ($\sqrt{counts}$) is the standard deviation associated with the measurement.

Each occurrence is statistically weighted to fit a Gaussian profile, whose standard deviation ($\sigma_T$) is the measure of total disorder. The standard error associated with the total disorder is shown in Figure 2 (D), (H), and (L) for the different interfaces and Figure 3 (A) and (B) for the different PM6 domains (neat and BHJ) of the main manuscript.

## 7. Weighted fits to the histograms obtained from PM6:Y6 and PM6:PC$_{71}$BM

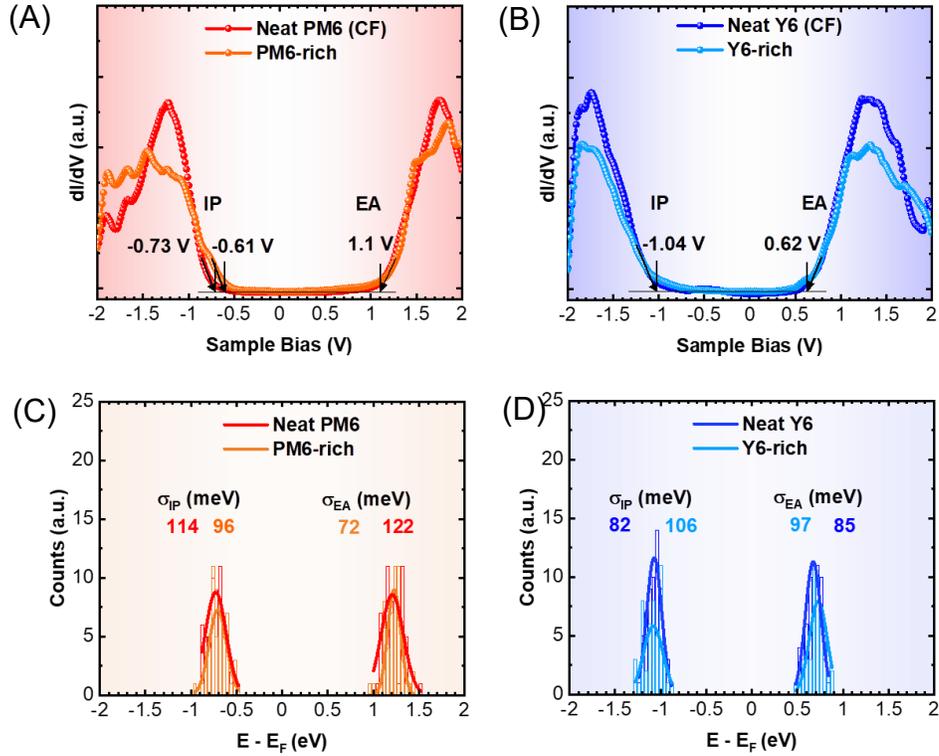

**Figure S10**: (A) & (B) Averaged dI/dV spectra of neat PM6 and Y6 as compared with the PM6 and Y6-rich phases in the PM6:Y6 blend, respectively. (C) & (D) Histogram obtained from the IP and EA onsets of neat PM6 and Y6 compared with the PM6 and Y6-rich phases in PM6:Y6 blend, respectively. The histograms were compiled from 50-60 dI/dV spectra acquired from different spatial positions in the neat and blend film corresponding to the PM6 and Y6-rich phases in the PM6:Y6 blend using 0.04 eV bins.

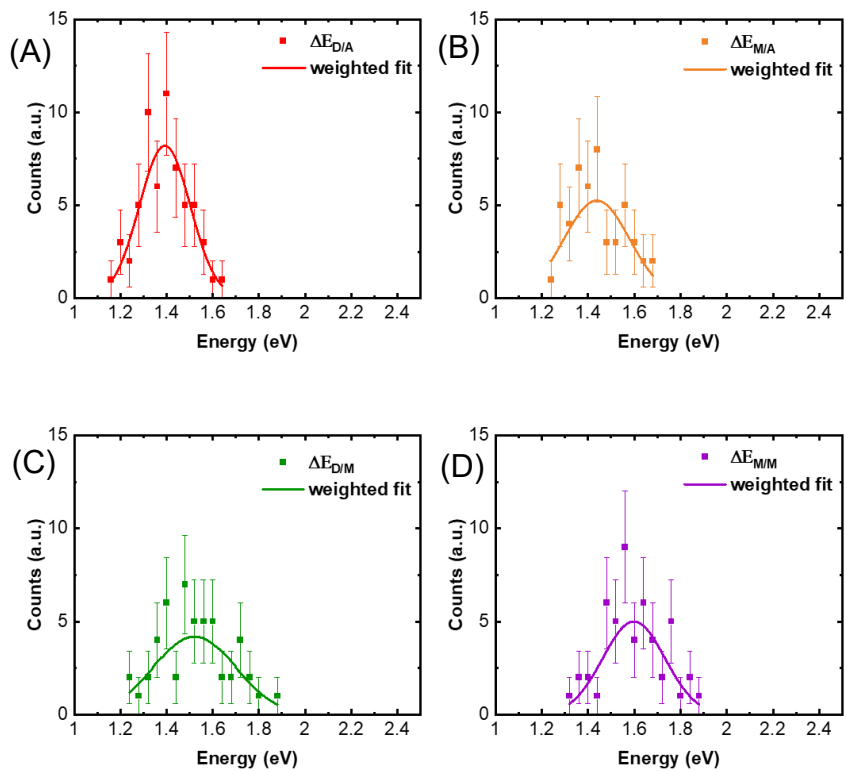

**Figure S11**: Weighted Gaussian fits to the histogram obtained from the energetic difference of the IP and EA from the (A) sharp D/A interfacial region, (B) M and Y6-rich domains, (C) PM6-rich and M domains, and (D) within the M domain, respectively. The data points are the number of occurrences using 0.04 eV energy bins with Poisson error bars.

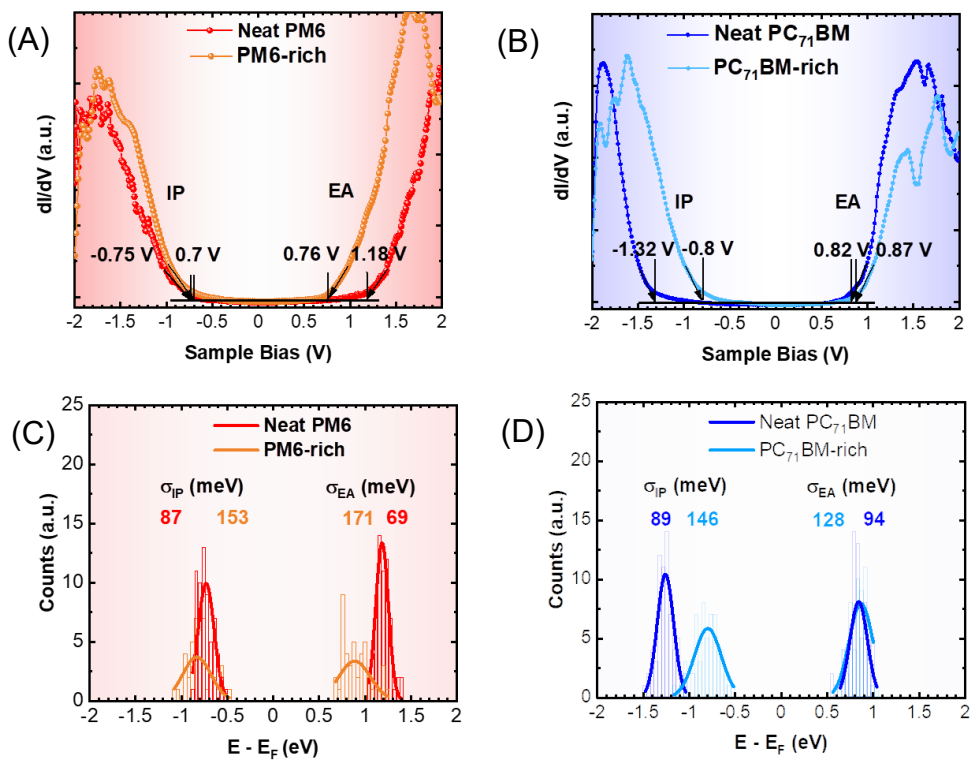

**Figure S12**: (A) & (B) Averaged dI/dV spectra of neat PM6 and PC$_{71}$BM as compared with the PM6 and PC$_{71}$BM-rich domains in the PM6:PC$_{71}$BM blend, respectively. (C) & (D) Histogram obtained from the IP and EA onsets of neat PM6 and PC$_{71}$BM compared with the PM6 and PC$_{71}$BM-rich domains in PM6:PC$_{71}$BM blend, respectively. The histograms were compiled from 50-60 dI/dV spectra acquired from different spatial positions in the neat and blend film corresponding to the PM6- and PC$_{71}$BM-rich domains in the PM6:PC$_{71}$BM blend using 0.04 eV bins.

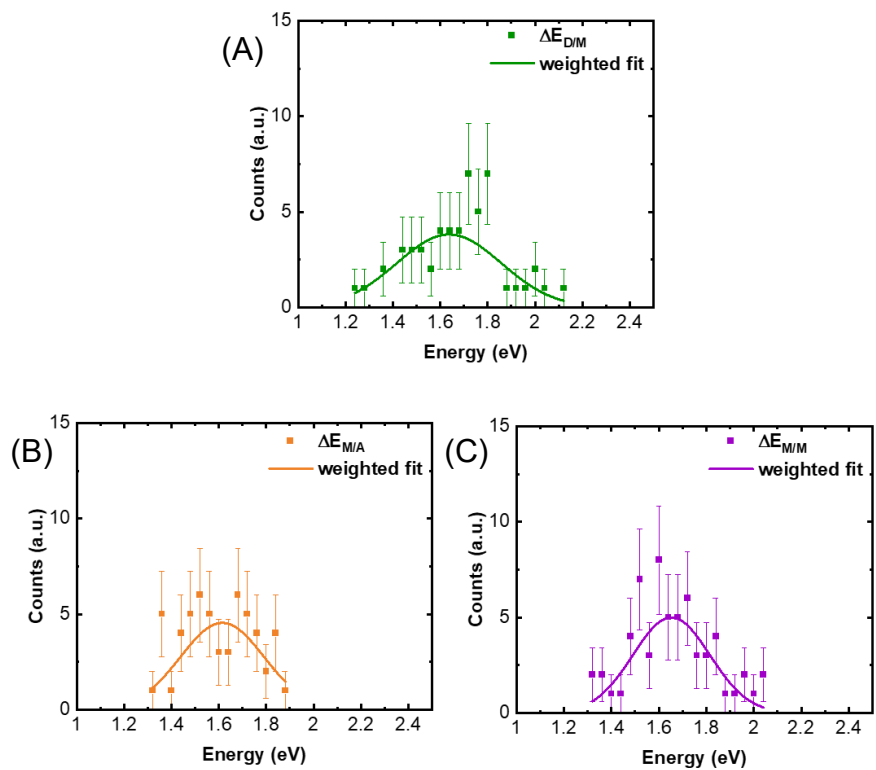

**Figure S13**: Weighted Gaussian fits to the histogram obtained from the energetic difference of the IP and EA from the (A) sharp D/M interfacial region (since D/A is absent), (B) M and PC$_{71}$BM-rich domains, (C) PM6-rich and M domains, and (D) within the M domain, respectively. The data points are the number of occurrences using 0.04 eV energy bins with Poisson error bars.

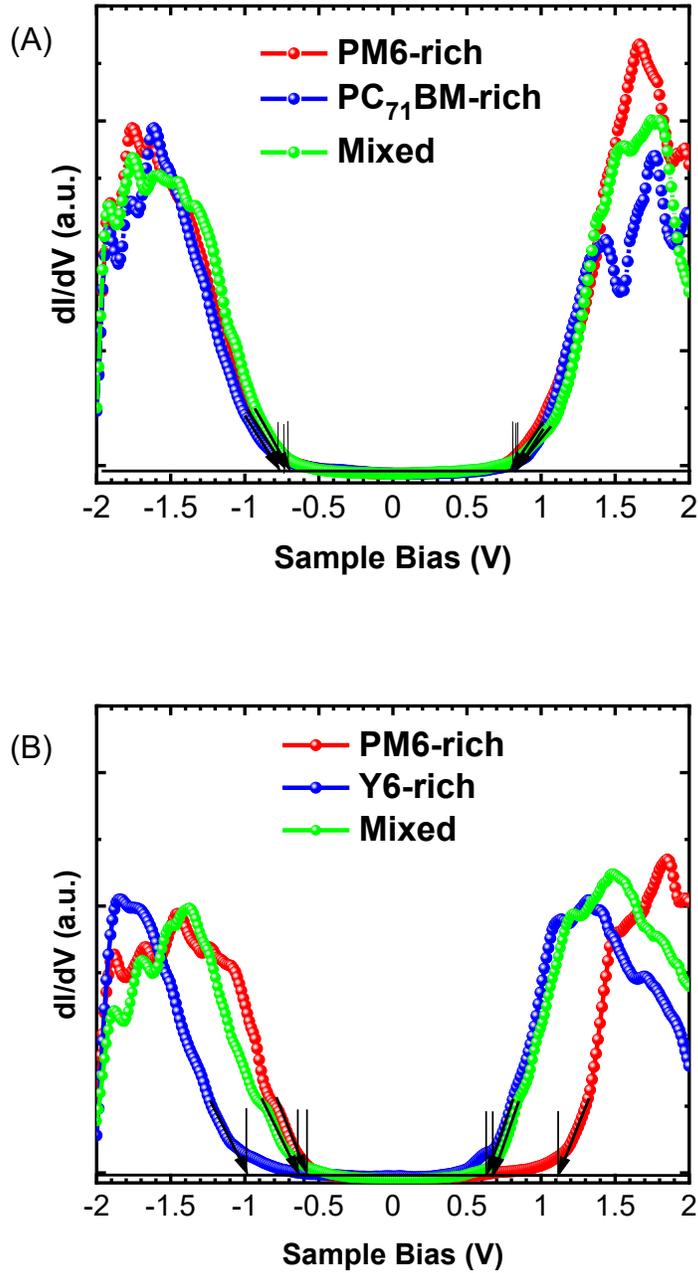

**Figure S14**: Averaged dI/dV spectra collected from the red, blue, and green areas of various dI/dV maps corresponding to the (A) PM6-rich, PC$_{71}$BM-rich and the M domains in the PM6: PC$_{71}$BM blend, (B) PM6-rich, Y6-rich and the M domains in the PM6:Y6 blend.

## 8. Fidelity between thin and thick BHJs and devices

Our study utilizes 10-15 nm thick BHJ layers for the purpose of STM/S and s-EQE measurements to probe the CT states manifold and energetic landscape. To compare the difference between thin and thick BHJs, we have fabricated thin and thick devices as mentioned in the device preparations section (Figure S15).

Not surprisingly, the thin BHJ devices exhibit reduced short-circuit current density ($J_{sc}$) owing to reduced absorption and light harvesting (Table S4). The thick and thin devices yield similar $V_{oc}$ values for PM6:IT4F and PM6:Y6 (CF) and similar s-EQE in the sub-band gap regions. Slight differences in $V_{oc}$ and s-EQE are observed in the cases of PM6:PC$_{71}$BM and PM6:Y6 (oDCB) where $V_{oc}$ differences of 50 mV and 30 mV are observed, respectively. Analysis of the non-radiative voltage loss for thin and thick devices shows a systematic difference (Tables S4 and S5). The difference in non-radiative voltage loss between thin and thick films was previously ascribed to optical interference effects as opposed to charge generation or transport phenomenon[10].

Importantly, to confirm that the underlying morphology and hence the disorder are functionally similar, we have calculated the static disorder from apparent Urbach energy analysis which shows that these values are similar within the error of the analysis for the thick and thin devices (Table S6). Moreover, we conducted AFM measurements that show similar morphology for the thick and thin samples (Figure S17).

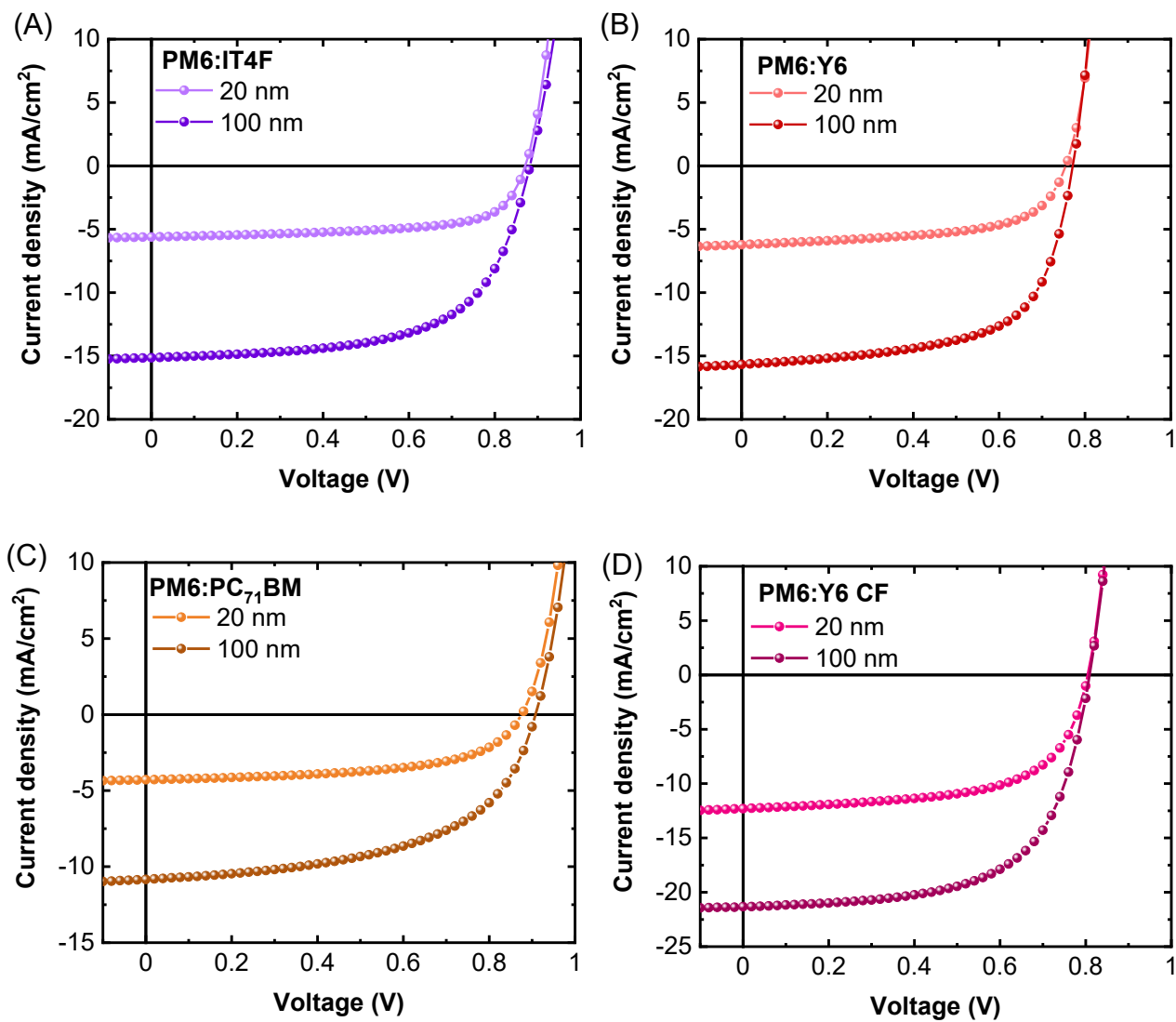

**Figure S15**: Comparison of J-V characteristics of thin and thick OPV devices of (A) PM6:IT4F, (B) PM6:Y6 (oDCB), (C) PM6:PC$_{71}$BM and (D) PM6:Y6 (CF). Small differences observed in V$_{OC}$ indicate the functional morphology of the thin active layer is nominally similar and representative of the thicker active layer morphology.

**Table S4:** Device performance parameters for the studied blends (processed with o-DCB) unless specified. Average values were obtained from 16 devices.

| Active Materials | $J_{SC}$ (mAcm$^{-2}$) | Jsc (sEQE) | $V_{OC}$ (V) | FF (%) | PCE (%) |
|---|---|---|---|---|---|
| PM6:PC$_{71}$BM (20 nm) | 4.43±0.18 | 4.5 | 0.874±0.003 | 57.33±0.37 | 2.22±0.09 |
| PM6:PC$_{71}$BM (100 nm) | 10.87±0.19 | 10.99 | 0.928±0.003 | 54.22±1.35 | 5.47±0.20 |
| PM6:IT4F (20 nm) | 5.44±0.17 | 3.14 | 0.867±0.005 | 64.71±2.48 | 3.06±0.22 |
| PM6:IT4F (100 nm) | 13.56±0.32 | 15.5 | 0.872±0.006 | 66.85±0.96 | 7.92±0.30 |
| PM6:Y6 (20 nm) | 6.35±0.34 | 6.47 | 0.737±0.010 | 59.75±1.02 | 2.80±0.15 |
| PM6:Y6 (100 nm) | 15.48±0.16 | 16.2 | 0.766±0.003 | 62.22±1.44 | 7.36±0.22 |
| PM6:Y6 (20 nm) (CF) | 11.44±0.73 | 10.46 | 0.8±0.010 | 61.94±1.14 | 5.75±0.51 |
| PM6:Y6 (100 nm) (CF) | 21.76±0.44 | 22.9 | 0.803±0.007 | 60.29±1.16 | 10.54±0.26 |
| PM6:Y6 (100 nm) (CF+0.5% v/v CN, 110°C for 5 mins) | 26.56±0.67 | 25.07 | 0.82±0.004 | 74.68±0.67 | 16.24±0.27 |

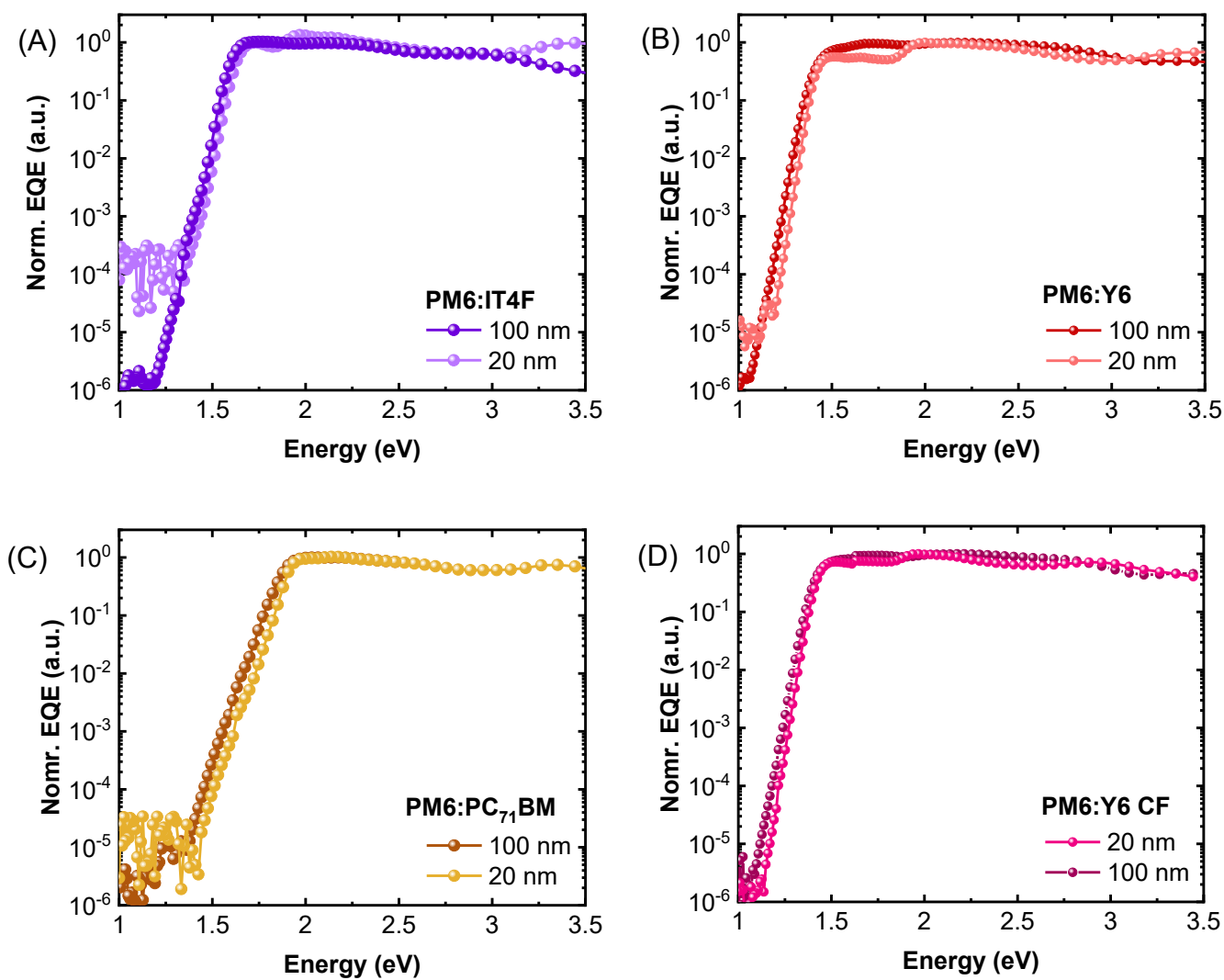

**Figure S16:** Normalized s-EQE of thin and thick (A) PM6:IT4F, (B) PM6: Y6 (oDCB), (C) PM6:PC71BM (D) PM6:Y6 (CF) OSCs.

**Determination of radiative and non-radiative recombination losses: ΔVr and ΔVnr:**

In OSCs, $V_{OC}$ is defined as[11]:

$$V_{OC} = \frac{n_i K_B T}{e} \ln\left(\frac{J_{PH}}{J_0} + 1\right) \qquad (7)$$

where $n_i$ is ideality factor, $J_{PH}$ is the photocurrent density in OPVs under open circuit voltage assumed to be equivalent to $J_{SC}$, $J_0 = J_{rad}$ represents the radiative recombination limits for the dark saturation current is determined form s-EQE as:

$$J_{rad} = e \int EQE(E) . \emptyset_{BB} . dE \qquad (8)$$

where $\emptyset_{BB}$ represents the black body radiation photon flux.

By incorporating the derived value of $J_{rad}$, radiative voltage ($V_r$) is determined using the following equation:

$$V_r = \frac{K_B T}{e} \ln\left(\frac{J_{PH}}{J_{rad}} + 1\right) \qquad (9)$$

Finally, the non-radiative recombination loss ($\Delta V_{nr}$) is then determined as,

$$\Delta V_{nr} = V_r - V_{OC} \qquad (10)$$

and $\Delta V_r$ is determined as,

$$\Delta V_r = \frac{E_{CT}}{e} - V_{OC} - \Delta V_{nr} \qquad (11)$$

The loss values determined from s-EQE studied OPVs in this work are listed below in Supplementary Table S5.

**Table S5:** Voltage loss parameters determine using s-EQE measurements for the OSCs. Total voltage loss calculated as $V_{loss} = E_{EQE,PV}/e - V_{OC}$

| Active Materials | $E_{EQE,PV}/e$ (V) | $E_{CT}/e$ (V) | $V_{OC}$ (V) | $\Delta V_r$ (V) | $\Delta V_{nr}$ (V) | $V_{loss}$ (V) |
|---|---|---|---|---|---|---|
| PM6:PC$_{71}$BM (o-DCB) | 1.89 | 1.65 | 0.928 | 0.264 | 0.40 | 0.962 |
| PM6:IT-4F (o-DCB) | 1.61 | 1.48 | 0.872 | 0.288 | 0.33 | 0.738 |
| PM6:Y6 (CF) | 1.42 | 1.39 | 0.803 | 0.333 | 0.25 | 0.617 |

**Table S6:** Comparison of the open circuit voltage ($V_{oc}$), non-radiative loss ($\Delta V_{nr}$) and the static disorder ($\sigma_s$) obtained from the apparent Urbach energy analysis for the thick and thin devices

| Active Materials | $V_{OC}$ (V) | $\Delta V_{nr}$ (V) | $\sigma_s\ (meV)$ using apparent Urbach analysis |
|---|---|---|---|
| PM6:PC$_{71}$BM (20 nm) | 0.874±0.003 | 0.46 | 62 ± 2.4 |
| PM6:PC$_{71}$BM (100 nm) | 0.928±0.003 | 0.4 | 65 ± 2.2 |
| PM6:IT-4F (20 nm) | 0.867±0.005 | 0.36 | 50.3 ± 2.9 |
| PM6:IT-4F (100 nm) | 0.872±0.006 | 0.33 | 51 ± 1.8 |
| PM6:Y6 (20 nm) | 0.8±0.010 | 0.28 | 47.5 ± 2.8 |
| PM6:Y6 (100 nm) | 0.803±0.007 | 0.25 | 48 ± 0.5 |

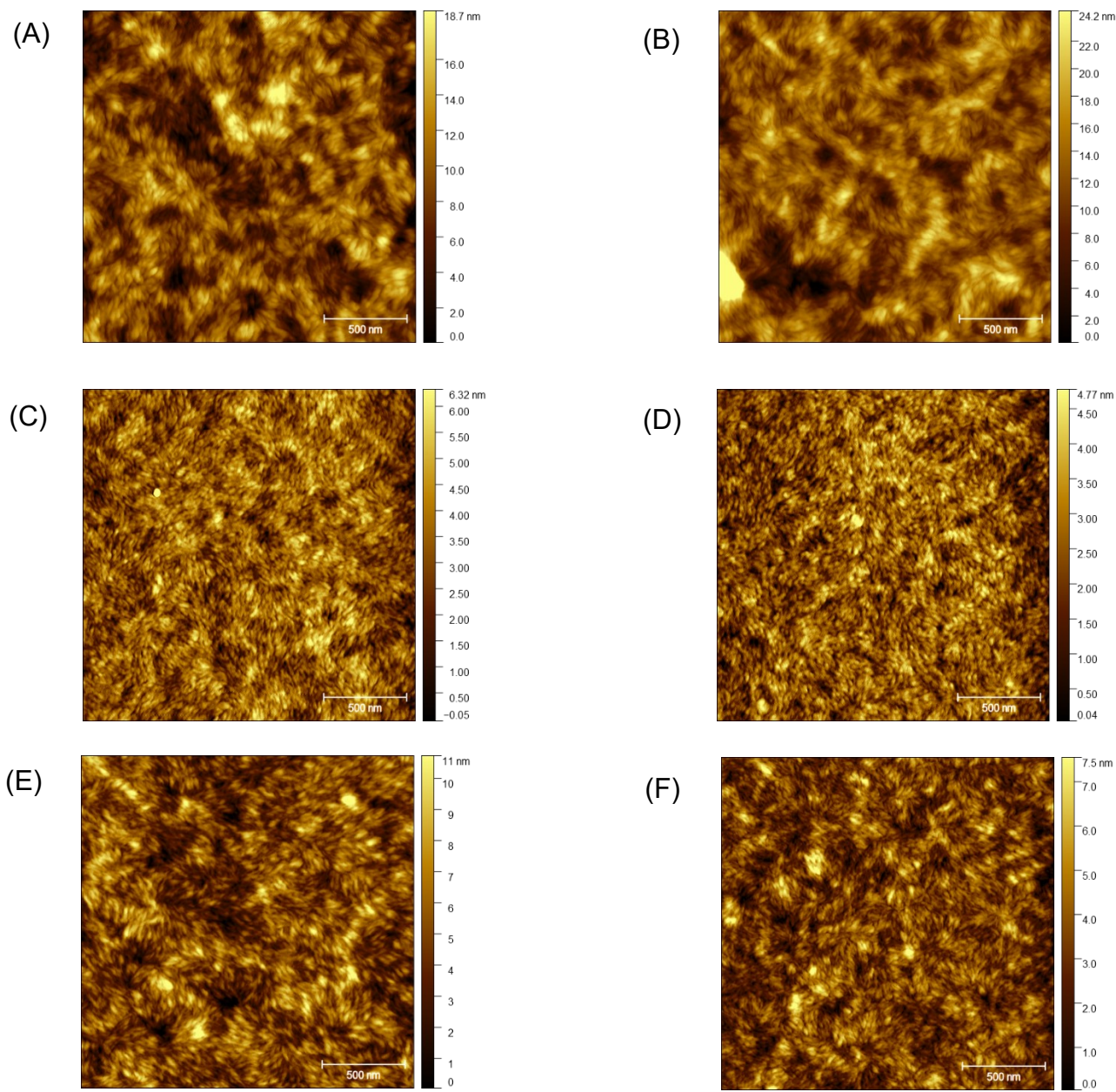

**Figure S17:** AFM topographical images of thick (A), (C) and (E) and thin (B), (D) and (F) samples of PM6:IT4F, PM6:Y6 (CF) and PM6:PC$_{71}$BM respectively. The surface roughness for the thick PM6:IT4F, PM6:Y6 (CF) and PM6:PC71BM are 3.14 nm, 0.92 nm, and 1.71 nm and thin PM6:IT4F, PM6:Y6 (CF) and PM6:PC71BM are 3.93 nm, 0.73 nm, and 1.14 nm respectively.

## 9. Absorbance spectra of neat PM6 and blend films using different solvents

The normalized absorbance spectra of neat and blend films prepared using the same protocol (Table S2) with different solvents shows the same trend in the level of PM6 aggregation in the blends using different solvents. It is obvious to see from Figure S10(C) that PM6:IT4F has a higher level of PM6 aggregation feature followed subsequently by PM6:Y6, neat PM6 and PM6:PC$_{71}$BM.

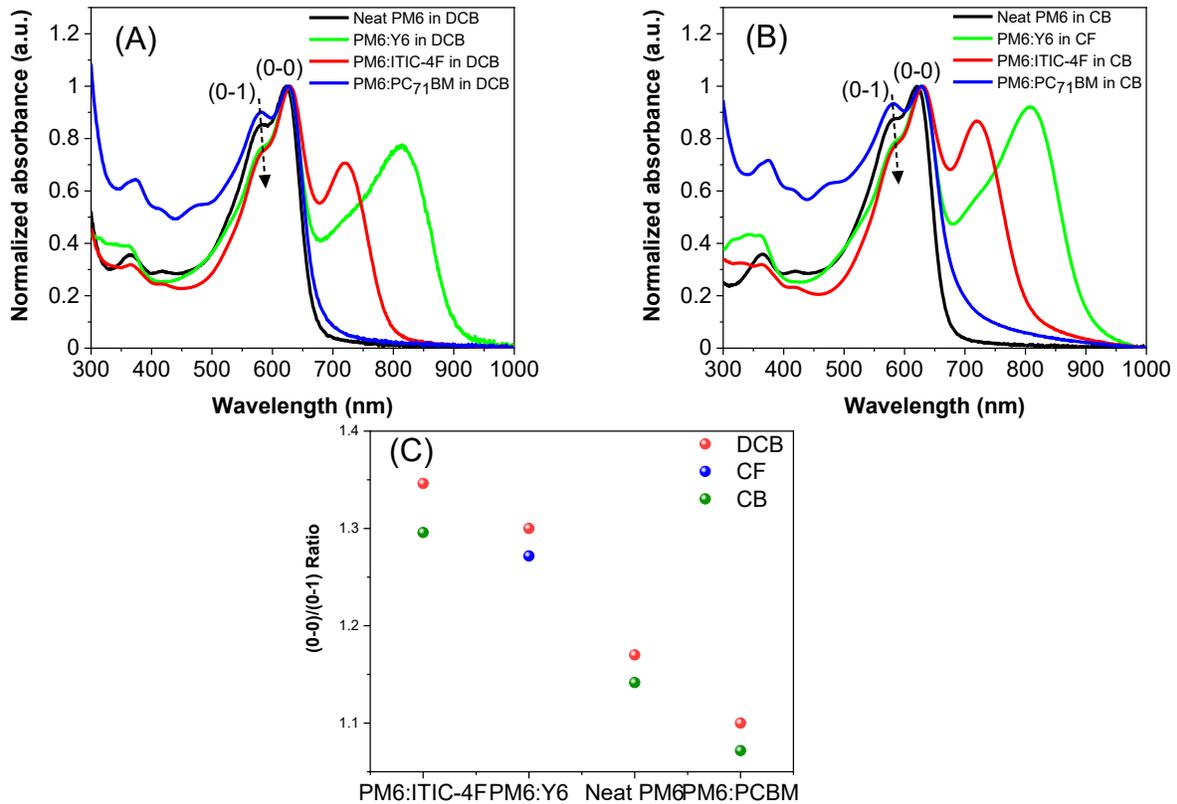

**Figure S18**: Normalized absorbance spectra of Neat PM6 and PM6:acceptor blend using (A) o-DCB as solvent and (B) CB as solvent for neat PM6, PM6:IT4F and PM6:PC$_{71}$BM and CF as solvent for PM6:Y6.

## 10. Line scans across sharp interface in PM6/Y6 interface

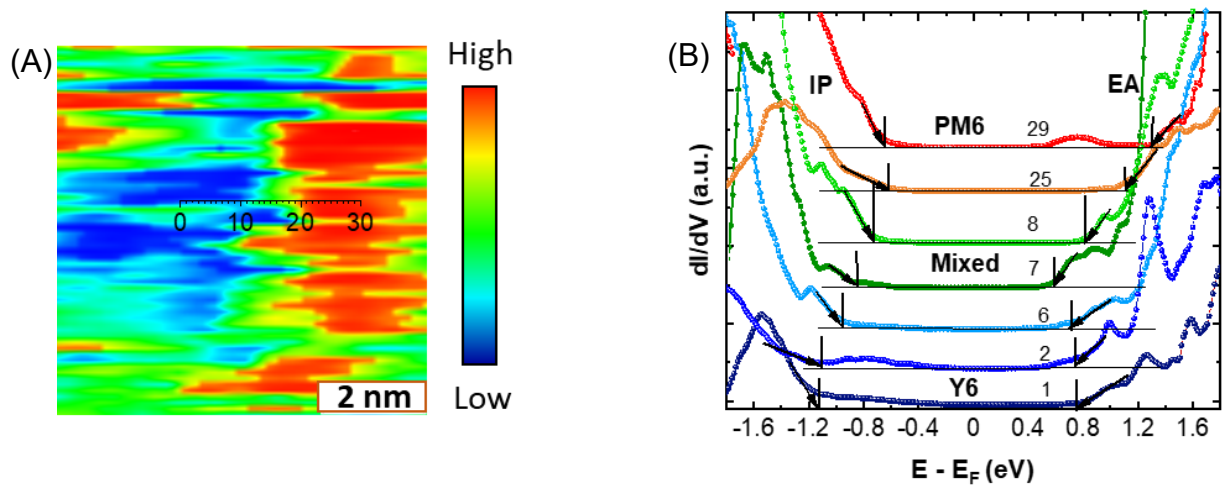

**Figure S19**: Magnified section of the dI/dV map of PM6:Y6 probed at (-1V, 500 pA) tunneling setpoints where spectroscopic measurement is performed along a line. The region covered in red, and blue are energetically identified as PM6- and Y6-rich domains, respectively, with green region associated with the interface with energetic offset as compared to the PM6 and Y6 spectra. (B) The dI/dV spectra obtained for selected spectra from the line spectroscopy measurement.

## 11. Determination of CT state properties & static disorder PM6:Y6 and PM6:PC$_{71}$BM

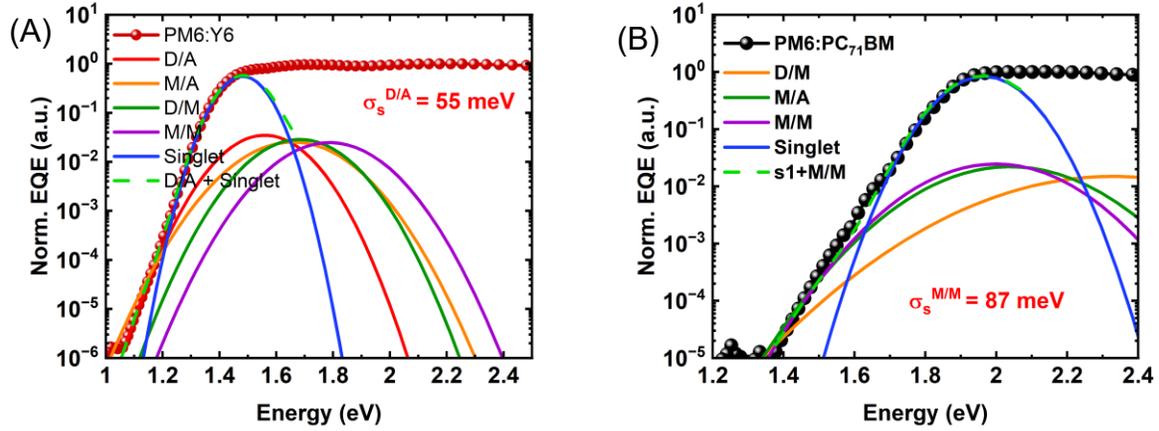

**Figure S20**: s-EQE from devices using (A) PM6:Y6 and (B) PM6:PC$_{71}$BM as an active layer with the CT states fit corresponding to the different interfaces in BHJ using the electronic distribution obtained from the STM/S analysis.

## 12. Determination of static disorder using STM/S & s-EQE for PM6:Y6(CF + CN)

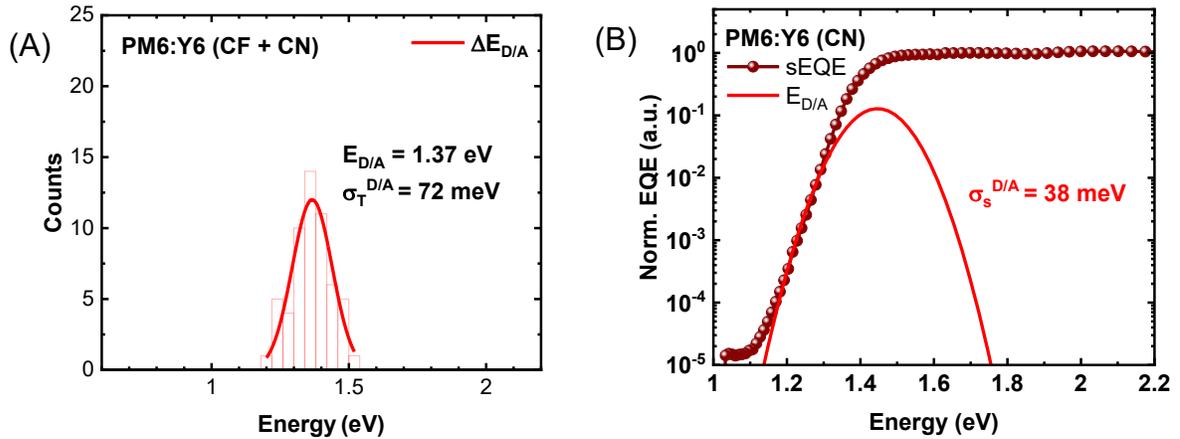

**Figure S21:** (A) Weighted fit to the histogram obtained from the sharp interfacial region of PM6:Y6(CF+CN). (B) s-EQE of the corresponding device along with the CT state fitting corresponding to $\Delta E_{D/A}$ using distribution obtained from STM/S measurements.

## 13. Device J-V curve of PM6:Y6(CF+CN) and apparent Urbach analysis

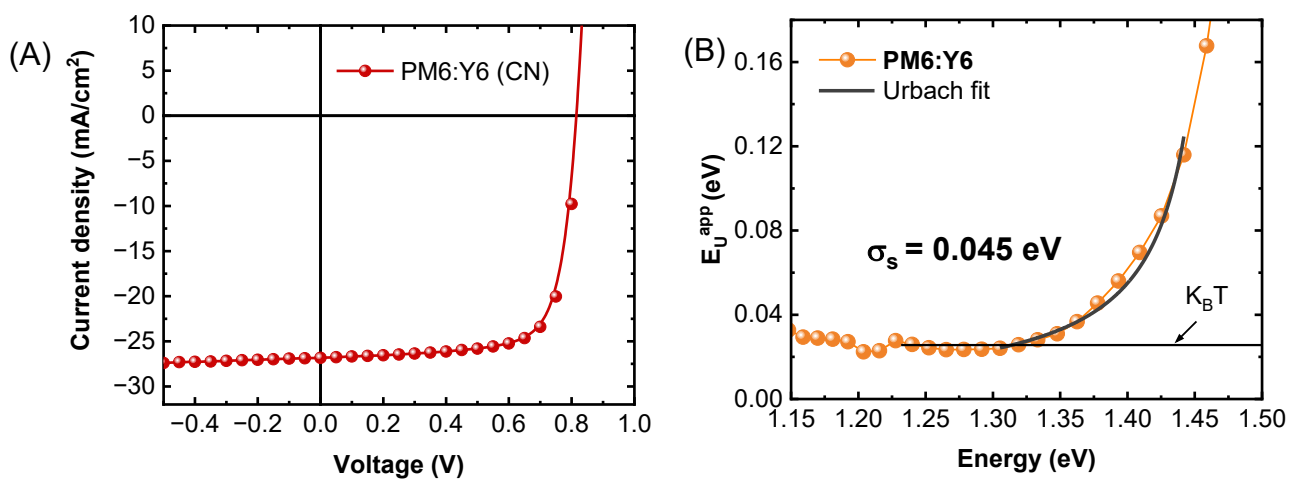

**Figure S22:** (A) J-V curve obtained from the PM6:Y6 device using 0.5% v/v CN as solvent additive and active layer thermally annealed at 110 °C for 5 mins. (B) Apparent Urbach energy and Urbach fit performed as explained for the blends studied (see **Approach 2**).

## 14. Determination of S1-CT offsets

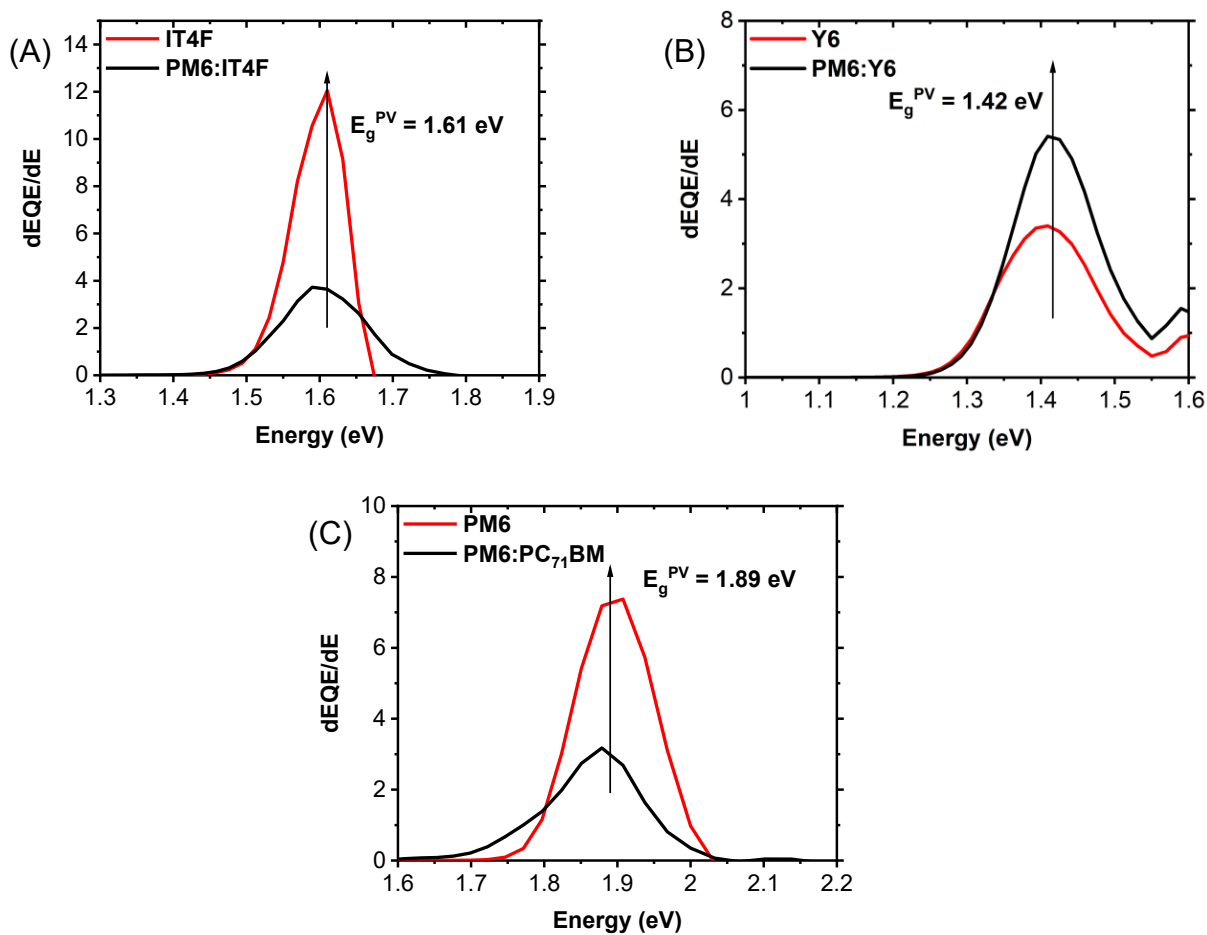

**Figure S23:** Derivative of s-EQE for (A) PM6:IT4F and neat IT4F, (B) PM6:Y6 and neat Y6 and (C) PM6:PC$_{71}$BM and neat PM6. The photovoltaic gap for the blend system and singlet energy gap for the neat devices are the same, from which we can get the singlet energy for the blend systems. Using the CT state energy as obtained from **Figure S9** and **Table S3**, we can obtain the S1- CT offset as (1.61 – 1.48 = 0.13 eV), (1.42 – 1.39 = 0.03 eV) and (1.89 – 1.65 = 0.24 eV) for PM6:PC$_{71}$BM, PM6:IT4F, and PM6:Y6 respectively.

## 15. Non-radiative voltage loss and apparent Urbach energy analysis for PM6:Y6 using different solvents and processing conditions

To further delineate the impact of interfacial disorder on the resulting non-radiative voltage loss, we fabricate PM6:Y6 devices using several different solvents and processing conditions. These cause changes in the underlying morphology[12] which is evident from the performance of these devices (**Figure S24**). These morphological changes can induce variation in many factors, like the electroluminescence efficiency, triplet exciton recombination fraction and so on, which can explain the higher efficiency and lower non-radiative voltage loss. Although a comprehensive understanding of all these factors is beyond the scope of this paper, an important focus remains on directly observing the influence of morphological changes on electronic disorder. Since we are dealing with the same system, we expect and observe negligible changes on the $\Delta E_{S1-CT}$ despite small shifts in the S1 and CT energies caused by the morphological variation (**Figure S24 and Table S7**). The S1-CT offset analysis for the additional PM6:Y6 oXY and oDCB has been performed using the traditional Marcus fit which is less accurate than the combined methodology that we have used for the PM6:Y6 CF and PM6:Y6 (CF+CN). However, the non-radiative voltage loss varies from 0.28 eV to 0.23 eV and the apparent Urbach energy analysis shows that the static disorder of the CT state has been reduced from 52 meV to 45 meV (**Table S6**). This observation highlights that non-radiative recombination is influenced by interfacial disorder even when the change in ΔE$_{S1\text{-}CT}$ is minimal.

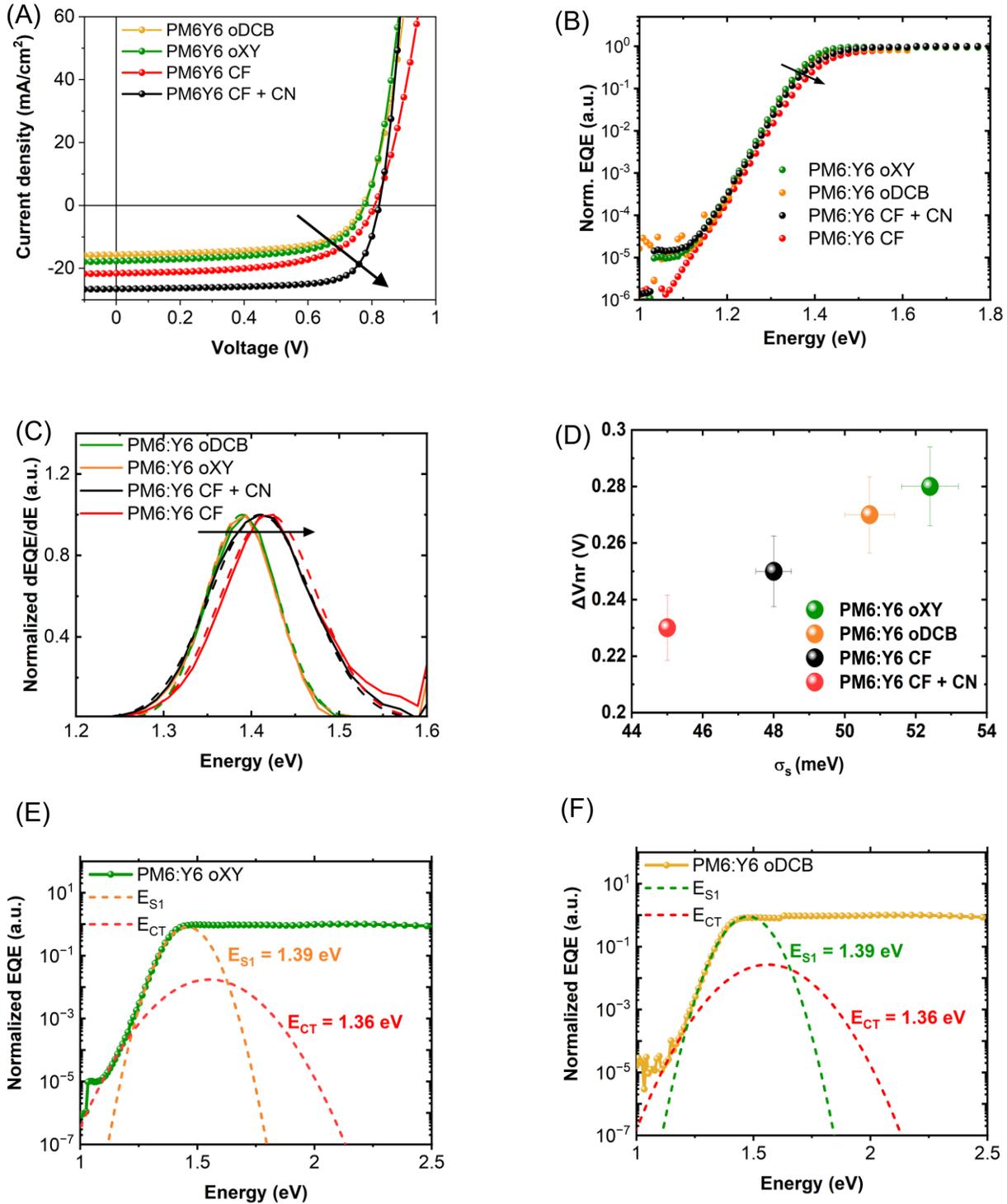

**Figure S24:** (A) J-V curve of PM6:Y6 using different solvents and processing conditions. (B) Normalized sEQE for PM6:Y6 using different solvents and processing conditions used in the calculation of the non-radiative voltage loss and static disorder using apparent Urbach energy analysis. (C)The normalized dEQE/dE for the mentioned blend systems. (D) Non-radiative voltage loss as a function of static disorder as obtained from the apparent Urbach energy analysis. sEQE fitting for the singlet and CT states for (E) PM6:Y6 oXY and (F) PM6:Y6 oDCB

**Table S7:** Non-radiative voltage loss and static disorder parameter extracted from the apparent Urbach analysis for PM6:Y6 using different solvent and processing conditions

| PM6:Y6 from solvents | $\Delta V_{nr}$ (V) | $E_{S1}$ (eV) | $E_{CT}$ (eV) | $\Delta E_{S1-CT}$ (eV) | $\sigma_s$ from apparent Urbach energy analysis (meV) |
|---|---|---|---|---|---|
| o-xylene | 0.28 | 1.39 | 1.36 | 0.03 | 52.4 ± 0.8 |
| oDCB | 0.27 | 1.39 | 1.36 | 0.03 | 50.7 ± 0.7 |
| CF + CN | 0.23 | 1.41 | 1.37 | 0.04 | 45 ± 0.2 |
| CF | 0.25 | 1.42 | 1.39 | 0.03 | 48 ± 0.5 |

12. Zhu, L., Zhang, M., Zhou, G., Hao, T., Xu, J., Wang, J., Qiu, C., Prine, N., Ali, J., Feng, W., et al. (2020). Efficient Organic Solar Cell with 16.88% Efficiency Enabled by Refined Acceptor Crystallization and Morphology with Improved Charge Transfer and Transport Properties. Adv. Energy Mater. *10*, 1904234. 10.1002/aenm.201904234.

13. Hosseini, S.M., Wilken, S., Sun, B., Huang, F., Jeong, S.Y., Woo, H.Y., Coropceanu, V., and Shoaee, S. (2023). Relationship between Energetic Disorder and Reduced Recombination of Free Carriers in Organic Solar Cells. Adv. Energy Mater. *13*, 2203576. 10.1002/aenm.202203576.

14. Perdigón-Toro, L., Zhang, H., Markina, A., Yuan, J., Hosseini, S.M., Wolff, C.M., Zuo, G., Stolterfoht, M., Zou, Y., Gao, F., et al. (2020). Barrierless Free Charge Generation in the High-Performance PM6:Y6 Bulk Heterojunction Non-Fullerene Solar Cell. Adv. Mater. *32*, 1906763. 10.1002/adma.201906763.
41